\documentclass[envcountsect,orivec]{llncs} 
\pdfoutput=1
\usepackage{etex} 
\usepackage[]{graphicx} 
\usepackage{times}
\usepackage[pdftex, colorlinks=true]{hyperref}
\newif\ifignore 
\ignorefalse
\newcommand{\auxproof}[1]{
\ifignore\mbox{}\newline
\textbf{BEGIN: AUX-PROOF} \dotfill\newline
{#1}\mbox{}\newline
\textbf{END: AUX-PROOF}\dotfill\newline
\fi}

\usepackage{verbatimbox}
\usepackage{amsthm}
\usepackage{fancybox,amssymb,amsfonts,amstext,amsmath,stmaryrd,wasysym,cite,mathtools,mathrsfs}
\usepackage{xspace} 
\allowdisplaybreaks[1] 

\usepackage[pdftex,all]{xy}
\CompileMatrices
\xyoption{v2}
\xyoption{curve}
\xyoption{2cell}
\SelectTips{cm}{}  
\UseAllTwocells
\SilentMatrices

\newdir{ >}{{}*!/-8pt/@{>}}  
\newdir{|>}{%
!/1.6pt/@{|}*:(1,-.2)@^{>}*:(1,+.2)@_{>}}
\newdir{pb}{:(1,-1)@^{|-}}
  \def\pb#1{\save[]+<20 pt,0 pt>:a(#1)\ar@{pb{}}[]\restore}

\usepackage{wrapfig}
\newtheorem{mythm}{Theorem}[section]
\newtheorem{mylem}[mythm]{Lemma}
\newtheorem{mycor}[mythm]{Corollary}
\newtheorem{myprop}[mythm]{Proposition}
\theoremstyle{definition}
\newtheorem{myexpl}[mythm]{Example}
\newtheorem{mydef}[mythm]{Definition}
\newtheorem{myrem}[mythm]{Remark}

\spnewtheorem*{myproof}{Proof}{\itshape}{\rmfamily}

 \def\myqed{\qed}

\newcommand{\hyper}[1]{{}^{*}{\kern-1.2pt}{#1}}
\newcommand{\dt}{\ensuremath{\mathtt{d\hspace{-.05em}t}}}
\newcommand{\While}{\textsc{While}}
\newcommand{\Whiledt}{\While^{\dt}}

\newcommand{\Hoare}{\textsc{Hoare}}
\newcommand{\Hoaredt}{\Hoare^{\dt}}

\newcommand{\B}{\mathbb{B}}
\newcommand{\N}{\mathbb{N}}
\newcommand{\R}{\mathbb{R}}

\newcommand{\Pow}{{\mathcal P}}

\newcommand{\lfp}{{\rm lfp}}

\newcommand{\U}{\mathbb{U}}

\newcommand{\Var}{\mathbf{Var}}

\newcommand{\AExp}{\mathbf{AExp}}
\newcommand{\BExp}{\mathbf{BExp}}
\newcommand{\Cmd}{\mathbf{Cmd}}

\newcommand{\aop}{\mathrel{\mathtt{aop}}}
\newcommand{\true}{\mathtt{true}}
\newcommand{\false}{\mathtt{false}}
\newcommand{\ttrue}{\text{t\hspace{-0.1em}t}}
\newcommand{\ffalse}{\text{ff}}

\newcommand{\SKIP}{\mathtt{skip}}
\newcommand{\IF}{\mathtt{if}}
\newcommand{\THEN}{\mathtt{then}}
\newcommand{\ELSE}{\mathtt{else}}
\newcommand{\WHILE}{\mathtt{while}}
\newcommand{\DO}{\mathtt{do}}

\newcommand{\hsigma}{\boldsymbol{\sigma}}
\newcommand{\St}{\mathbf{St}}

\newcommand{\hS}{\mathbf{S}}

\newcommand{\Galois}[4]{#1 \overset{#4}{\underset{#3}\rightleftharpoons}
#2}

\newcommand{\Preord}{\mathsf{Preord}}
\newcommand{\Poset}{\mathsf{Poset}}
\newcommand{\Concr}{\mathsf{Concr}}
\newcommand{\AscCn}{\mathsf{AscCn}}
\newcommand{\Cpo}{\mathsf{Cpo}}
\newcommand{\Monotone}{\mathsf{Monotone}}
\newcommand{\Conti}{\mathsf{Conti}}
\newcommand{\Basis}{\mathsf{Basis}}
\newcommand{\Cover}{\mathsf{Cover}}
\newcommand{\Term}{\mathsf{Term}}
\newcommand{\Widen}{\mathsf{Widen}}
\newcommand{\WidenSeq}{\mathsf{WidenSeq}}
\newcommand{\UnifTerm}{\mathsf{UnifTerm}}
\newcommand{\UnifWiden}{\mathsf{UnifWiden}}

\newcommand{\CP}{\mathbb{CP}}

\newcommand{\con}{{\sf con}}
\newcommand{\gen}{{\sf gen}}
\newcommand{\Constr}{{\sf Constr}}

\newcommand{\basis}{{\ooalign{{$\bot$}\crcr{\hss{-}\hss}}}}

\newcommand{\sem}[1]{\llbracket #1 \rrbracket}

\newcommand{\semcp}[1]{\llbracket #1 \rrbracket_{\CP}}

\newcommand{\filt}{\mathcal{F}}
\newcommand{\pow}{\mathcal P}
\newcommand{\C}{\mathbb{C}}

\newcommand{\place}{\underline{\phantom{n}}\,} 
\newcommand{\rtuple}[1]{(#1)}  
\newcommand{\co}{\mathbin{\circ}}

\newcommand{\iso}{\mathrel{\stackrel{
           \raisebox{.5ex}{$\scriptstyle\cong\,$}}{
           \raisebox{0ex}[0ex][0ex]{$\rightarrow$}}}}
\newcommand{\LU}{{\mathscr L}_{\R}}
\newcommand{\LUpr}{{\mathscr L}_{\hyper{\R}}}
\newcommand{\baseSet}{X}
\newcommand{\LX}{\mathscr{L}_{\baseSet}}
\newcommand{\LsX}{\mathscr{L}_{\hyper{\baseSet}}}

\newcommand{\CHypSt}{\hyper{(\C^{\infty})}}

\newcommand{\cto}{\to_{\mathrm{ct}}} 
 
\newcommand{\scto}{\to_{\hyper{\mathrm{ct}}}} 
\newcommand{\dsup}{\bigsqcup} 

\newcommand{\LR}{\mathscr{L}_{\R}}
\newcommand{\LsR}{\mathscr{L}_{\hyper{\R}}}
\newcommand{\sect}[1]{|_{#1}}

\title{Abstract Interpretation with Infinitesimals
\thanks{
We thank 
 Kohei Suenaga and the anonymous referees 
for useful discussions and comments.
  This research was supported in part by Grants-in-Aid No.\ 24680001 \& 15KT0012,  JSPS; Grant-in-Aid for JSPS Fellows; NSF CAREER award \#1156059; and NSF award \#1162076.
}
\\
{\large Towards Scalability in \emph{Nonstandard Static Analysis}}
}
\author{
  Kengo Kido \inst{1, 2}
  \and
  Swarat Chaudhuri \inst{3}
  \and
  Ichiro Hasuo \inst{1}
}
\institute{
  University of Tokyo, Japan
  \and
  JSPS Research Fellow
  \and
  Rice University, USA
  }

\begin{document}

\maketitle

  \begin{abstract}
We extend abstract interpretation for the purpose of verifying hybrid systems.
Abstraction has been playing an important role in many verification methodologies for hybrid systems, but
some special care is needed for abstraction of continuous dynamics defined by ODEs.
We apply Cousot and Cousot's framework of abstract interpretation to
   hybrid systems, almost \emph{as it is}, by regarding continuous
   dynamics as an infinite iteration of \emph{infinitesimal} discrete jumps.
This extension follows the recent line of work by Suenaga, Hasuo and Sekine, where deductive verification is extended for hybrid systems by 1) introducing a constant $\dt$ for an
  infinitesimal value; and 2) employing Robinson's \emph{nonstandard
  analysis (NSA)} to define mathematically rigorous semantics.
 Our theoretical results include soundness and termination via
 \emph{uniform} widening operators; and our prototype implementation
 successfully verifies some benchmark examples.

  \end{abstract}

\nocite{NSAI}

\section{Introduction}\label{sec:introduction}
\emph{Hybrid systems}  exhibit both discrete \emph{jump}
 and continuous  \emph{flow} dynamics. Quality assurance of such systems
 are of paramount importance due to the current ubiquity of
 \emph{cyber-physical systems (CPS)} like cars, airplanes, and many
 others.  
 For the formal verification approach to hybrid systems, the challenges are: 1) to incorporate
 flow-dynamics; and 2) to do so at the lowest possible cost, so that the
 existing discrete framework smoothly transfers to hybrid situations.  A large
 body of existing work uses \emph{differential equations} explicitly in the
 syntax; see the discussion of related work below.

In~\cite{Suenaga2011}, instead, an alternative approach of
\emph{nonstandard static analysis}---combining \emph{static analysis}
and \emph{nonstandard analysis}---is proposed. Its basic idea is
to introduce a constant
$\dt$ for an \emph{infinitesimal} (i.e.\ infinitely small) value, and
\emph{turn flow into jump}. With $\dt$, the continuous operation of
integration can be represented by a while-loop, to which
existing discrete techniques such as Hoare-style program logics readily
apply. For a rigorous mathematical development they employ
\emph{nonstandard analysis (NSA)} beautifully formalized by
Robinson~\cite{Robinson1966}. 

Concretely, in~\cite{Suenaga2011} they took the common combination of a
$\textsc{While}$-language and a Hoare
logic (e.g.\ in the textbook~\cite{Winskel1993}); and added a constant $\dt$
to obtain a modeling and verification framework for hybrid systems.  Its
 components are called $\Whiledt$ and $\Hoaredt$. 
The soundness of $\Hoaredt$ is proved
against denotational semantics defined in the language
of NSA.
Subsequently in the \emph{nonstandard static analysis} program: in~\cite{Hasuo2012} they presented a prototype
automatic theorem prover for $\Hoaredt$;
and in~\cite{Suenaga2013} 
they applied the same idea to stream processing systems, realizing 
a verification framework for \emph{signal processing} as in Simulink.

  Underlying these technical developments is the idea of so-called \emph{sectionwise execution}.
Although this paper does not rely explicitly on it, it is still useful
for laying out the ``operational'' intuition of nonstandard static analysis.
See the following example.
\vspace*{1em}

\noindent
\begin{minipage}{.77\textwidth}
\begin{myexpl}\label{example:elapsedTime}
  Let $c_{\mathsf{elapse}}$ be the 
 program on the right.
 The value of $\dt$ is infinitesimal; therefore the
 $\mathtt{while}$ loop will not terminate within finitely many steps.
 Nevertheless it is somehow intuitive to expect that after an ``execution'' of
 this program, the value of $t$
 should be infinitesimally close to $1$ and larger than it.  
\end{myexpl}
\end{minipage}
\hfill
\begin{math}
 \begin{array}{l}
 t := 0\;;\quad
\\
 \mathtt{while}\; t\le 1\; \mathtt{do}
\\
 \quad t:=t+\dt
 \end{array}
\end{math}


\begin{wrapfigure}{r}{7em}
\begin{math}
 \begin{array}{l}
 t := 0\;;\quad
\\
 \mathtt{while}\; t\le 1\; \mathtt{do}
\\
 \quad t:=t+\frac{1}{i+1}
 \end{array}
\end{math}
\end{wrapfigure}

 One possible way of thinking is to imagine \emph{sectionwise execution}.  For each
 natural number $i$ we consider the \emph{$i$-th section} of the program
 $c_{\mathsf{elapse}}$, denoted by $c_{\mathsf{elapse}}\sect{i}$
 and shown on the right.
 Concretely, $c_{\mathsf{elapse}}\sect{i}$ is obtained by replacing the
 infinitesimal $\dt$ in $c_{\mathsf{elapse}}$ with $\frac{1}{i+1}$.
 Informally $c_{\mathsf{elapse}}\sect{i}$ is the ``$i$-th approximation'' of
 the original $c_{\mathsf{elapse}}$.

 A section $c_{\mathsf{elapse}}\sect{i}$ does terminate within
 finite steps and
 yields $1+\frac{1}{i+1}$ as the value of $t$.
 Now we collect the outcomes of sectionwise executions and obtain
 a sequence 
 \begin{equation}\label{equation:sequenceThatIsOnePlusDt}
 \small
 \begin{array}{l}
  (\,1+1, \;
 1+\frac{1}{2},\;
 1+\frac{1}{3},\;
 \dotsc,\;
 1+\frac{1}{i},\;
 \dotsc\,
 )
 \end{array}
 \end{equation}
 which is thought of as a progressive approximation of the actual
 outcome of the original program $c_{\mathsf{elapse}}$.  Indeed, in the
 language of  NSA, the  sequence~(\ref{equation:sequenceThatIsOnePlusDt}) represents a \emph{hyperreal number}
 $r$ that is infinitesimally close to $1$. 

\vspace*{.0em}

We note that
 a program in $\Whiledt$ is \emph{not} intended to be executed: the program $c_{\mathsf{elapse}}$
does not terminate. 
It is however an advantage of
\emph{static} approaches to verification and analysis,  that programs need not be executed to prove their
correctness. Instead well-defined mathematical semantics suffices. This
is what we do here as well as in~\cite{Suenaga2011,Hasuo2012,Suenaga2013}, with the denotational
semantics of $\Whiledt$ exemplified in Example~\ref{example:elapsedTime}.

\vspace*{.2em} 
\noindent
\textbf{Our Contribution}
\quad
In the previous work~\cite{Suenaga2011,Hasuo2012,Suenaga2013} \emph{invariant
discovery} has been a big obstacle in scalability of the proposed
verification techniques---as
is usual in deductive verification. The current work, as a first step
towards scalability of the approach,
extends \emph{abstract interpretation}~\cite{Cousot1977} with
infinitesimals. The  abstract interpretation methodology is known for
its ample applicability (it is employed in model checking as
well as in many deductive verification frameworks) and scalability (the static
analyzer Astr\'{e}e~\cite{Cousot2005} has been successfully used e.g.\
for Airbus's flight control system).

Our theoretical contribution includes: the theory of \emph{nonstandard
abstract interpretation} where (standard) abstract domains are ``$*$-transformed,'' in a rigorous NSA sense, to 
the abstract domains for hyperreals; their soundness in over-approximating
semantics of $\Whiledt$ programs and hybrid system modeling by them; and introduction of the notion of
\emph{uniform} widening operators. With the latter, inductive
approximation is guaranteed to terminate within finitely many
steps---even after extension to the nonstandard setting. We show that
many known widening operators, if not all, are indeed uniform.
Although we  focus on the domain of convex polyhedra in this paper, it
is also possible to extend other abstract domains like
ellipsoids~\cite{Feret2004} 
in the same way.

These theoretical results form a basis of our prototype
implementation,\footnote{The prototype is available on-line: \href{http://www-mmm.is.s.u-tokyo.ac.jp/~kkido/}{http://www-mmm.is.s.u-tokyo.ac.jp/\~{}kkido/}} that successfully analyzes: \emph{water-level monitor},
a common example of piecewise-linear hybrid dynamics; and also
\emph{thermostat} that is beyond piecewise-linear.
The prototype deals with the constant $\dt$ as a truly infinitesimal number using computer algebra system.

\vspace*{.2em}
\noindent
\textbf{Related Work}
\quad
%
%
%
%
There has been a lot of research work for verification of hybrid systems
and it 
has led to quite a few system verification tools, including
 HyTech~\cite{Henzinger1997},
 PHAVer~\cite{Frehse05},
 SpaceEx~\cite{Frehse11},
 HySAT/iSAT~\cite{Franzle2007},
 Flow*~\cite{ChenAS13} and
 KeYmaera~\cite{PlatzerQ08}. All these rely on ODEs (or the explicit
 solutions of them) for expressing continuous dynamics, much like 
\emph{hybrid automata}~\cite{Alur1992} do.

Our nonstandard static analysis approach is completely different from
those in the following point: we do not use ODEs at all, and model
hybrid systems as an imperative program with an infinitesimal constant.
It enables us to apply static methodologies for discrete systems as they are.  For example, in HyTech and PHAVer, convex polyhedra is used to
over-approximate the reachable sets.  They need, however, some special
techniques such as linear phase-portrait~\cite{Henzinger95}, to reduce the dynamics into
piecewise linear one.  Our framework does not need such and usual
abstract interpretation works as it is.

There are many other works we rely on, such as
those on
abstract interpretation, nonstandard analysis, etc. These are discussed 
later when they become relevant.

\vspace*{.2em}
\noindent
\textbf{Organization}
\quad
In~\S{}\ref{sec:exampleOfAnalysis} we start with the water-level monitor example and present how our nonstandard abstract interpretation framework works. 
Then we go on to its theoretical foundations.
In~\S{}\ref{sec:preliminaries} we review
preliminaries on: abstract
interpretation; nonstandard analysis; and the modeling language $\Whiledt$
from~\cite{Suenaga2011}. 
In~\S{}\ref{sec:NSAI} we extend the theory of abstract
interpretation with infinitesimals and build the theory of nonstandard
abstract
interpretation. 
Its theorems include soundness of approximation, and
 termination guaranteed by (the $*$-transform of) a \emph{uniform} widening
 operator. 
In~\S{}\ref{sec:implementation}
we present our prototype implementation and the experiment results with it.

Most proofs are deferred to Appendix~\ref{appendix:omittedProofs}.

\section{Leading Example: Verification of Water-Level Monitor}\label{sec:exampleOfAnalysis}
We shall start with an example of verification and let it exemplify how our
 framework---that extends abstract interpretation with infinitesimals,
 and handles continuous as well as discrete dynamics---works. We use the
 well-known example of the water-level monitor~\cite{Alur1992}.
 In the current section, in particular, we will first revisit how the usual
 abstract interpretation workflow (without extension) would work, using
 a 
 discretized variant of the problem. Our emphasis is on the fact that
 our extended framework works just in the same manner: without any
 explicit ODEs or any additional theoretical infrastructure for ODEs;
 but only adding a constant $\dt$.

\begin{wrapfigure}[4]{r}{0pt}
  \raisebox{-13mm}[0pt][0mm]{\includegraphics[width=.18\textwidth]{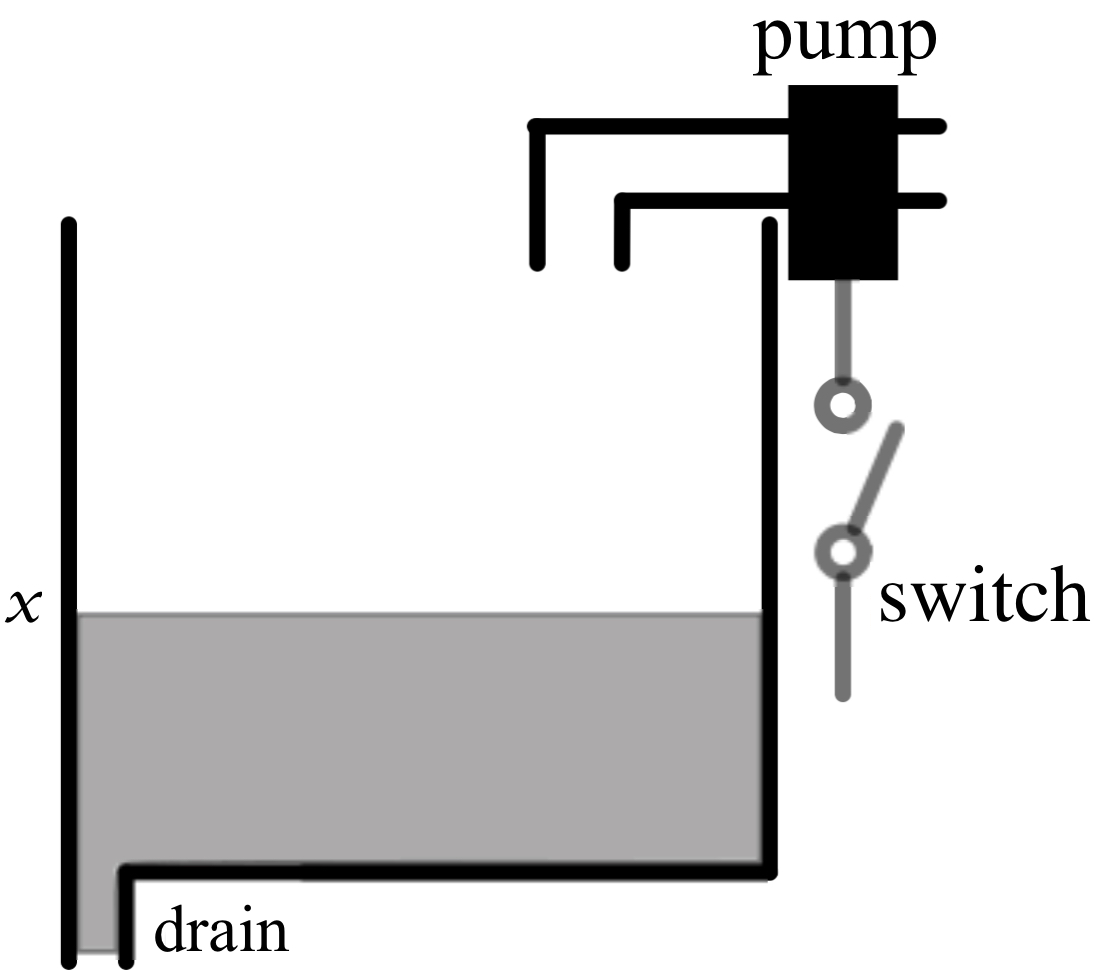}}
\end{wrapfigure}
 The concrete problem is as follows.
 See the figure on the right.  A water tank has a
 constant drain ($2$~cm per second). When the water level $x$ gets lower
 than $5$~cm the 
 switch is turned on, which eventually makes the pump work but only after 
 a time lag of two seconds. While the pump is working, the water level
 $x$
 rises by $1$~cm per second. Once $x$ reaches $10$~cm the switch is
 turned off, which will shut down the pump but again after a time lag of two seconds.
 Our goal is the \emph{reachability analysis} of this hybrid dynamics,
 that is, to see the water level $x$ remains in a certain ``safe'' range (we will see
 that the range is $1\le x \le 12$).



\begin{wrapfigure}[13]{r}{0.5\hsize}
\vspace{-1em}
 \begingroup
 \fontsize{7pt}{8pt}\selectfont
  \begin{verbatim}
(*Water-Level Monitor*)
l := 0; x := 1; p := 1; s := 0; 
dt' := 0.2;  
while true do {
   if p = 1 then x := x + dt' 
      else x := x - 2 * dt';
   if (x <= 5 && p = 0) then s := 1 
      else {if (x >= 10 && p = 1) 
               then s := 1 
               else s := 0
      };
   if s = 1 then l := l + dt'
      else skip;
   if s = 1 && l >= 2
      then {p := 1 - p; s := 0; l := 0}
      else skip
}
\end{verbatim}
\endgroup 
\vspace{-1.7em}
\caption{Discretized water-level monitor}
\label{fig:whileCodeCaseStudy}
\end{wrapfigure}
\subsection{Analysis by (Standard) Abstract Interpretation, as a Precursor}\label{subsec:waterLevel0.2}
Let us first revisit the usual workflow in reachability analyses by
abstract interpretation.
 We will use the \emph{discretized} model of the water-level monitor
 in Fig.~\ref{fig:whileCodeCaseStudy}, where each iteration of its unique
loop amounts to the lapse of $\dt'=0.2$ seconds. The model in
Fig.~\ref{fig:whileCodeCaseStudy} is 
an imperative program
with while loops, a typical subject of analyses by abstract
interpretation. 

More specifically:
 $x$ is the water level, $l$ is the counter for the time lag,
$p$ stands for  the state of the pump 
 ($p=0$ if the pump is off, and $p=1$ if on) 
and $s$ is for ``signals,'' 
meaning $s=1$ if the pump has not yet responded to a signal from the
 switch (such as, when the switch is on but the pump is not on yet). 


The first step in the usual abstract interpretation workflow is 
to fix \emph{concrete} and \emph{abstract domains}. Here in~\S\ref{subsec:waterLevel0.2} we will use the followings.
\begin{itemize}
 \item \textbf{The concrete domain: $\bigl(\pow({\R}^{2})\bigr)^{4}$.} 
       We have two numerical variables $l,x$ and two Boolean ones $p,s$
       in Fig.~\ref{fig:whileCodeCaseStudy},
       therefore a canonical concrete domain would be $\pow(\B^{2}\times
       \R^{2})$. We have the powerset operation $\pow$ in it since we
       are now interested in the \emph{reachable} set of memory states.

       However, for a better fit with our abstract domain  (namely
       convex polyhedra), we shall use the set
       $\bigl(\pow({\R}^{2})\bigr)^{4}$ that is isomorphic to the above set
$\pow(\B^{2}\times
       \R^{2})$.

 \item \textbf{The abstract domain: $(\CP_{2})^{4}$.} 
 We use the domain of \emph{convex polyhedra}~\cite{Cousot1978},
       one of the most commonly-used abstract domains. 
       Recall that a convex polyhedron is a subset of a Euclidean space
       characterized by a finite conjunction of linear inequalities. 
       Specifically, we
       let $\CP_{2}$, the set of 2-dimensional convex polyhedra, 
       approximate the set $\pow(\R^{2})$. Therefore, as an abstract
       domain for the program in Fig.~\ref{fig:whileCodeCaseStudy}, 
       we 
       take $(\CP_{2})^{4}$ (that approximates
       $\bigl(\pow({\R}^{2})\bigr)^{4}$). 

\end{itemize}


The next step in the workflow is to \emph{over-approximate} the set of
memory states that are reachable by the program in
Fig.~\ref{fig:whileCodeCaseStudy}---this  is a subset of the concrete domain
$\bigl(\pow({\R}^{2})\bigr)^{4}$---using the abstract domain
$(\CP_{2})^{4}$. Since the desired set can be thought of as a least
fixed point, this over-approximation procedure involves: 1)
\emph{abstract execution} of the program in $(\CP_{2})^{4}$ (that is
straightforward, see e.g.~\cite{Cousot1978}); and 2) acceleration of
least fixed-point computation in $(\CP_{2})^{4}$ via suitable use of a
\emph{widening operator}. For convex polyhedra  several
 widening operators are well-known. We shall use here $\nabla_M$, so-called the
\emph{widening up
to $M$} operator from \cite{Halbwachs1993, Halbwachs1997}. One big
reason for this choice is the \emph{uniformity} of the operator (a
notion we introduce later in~\S{}\ref{subsec:uniformWidening}), among
others. The set $M$ of linear constraints is a parameter for this
widening operator; we fix it as usual, collecting the linear constraints
that occur in the program in question. That is, $M=\{x\leq 5, x\geq 5,
x\leq 10, x\geq 10, l\leq 2, l\geq 2\}$. 

This over-approximation procedure is depicted in the \emph{iteration
sequence} in Fig.~\ref{fig:iterseq}. Let us look at some of its
details. The graph $0$ represents the initial memory state (before
the first iteration), where the pump is on and the water level $x$ is precisely
$1$. After one iteration the water level will be incremented by $1\times
\dt'=0.2$~cm; as usual in abstract interpretation, however, at this moment we invoke the widening
operator
$\nabla_M$, and the next ``abstract reachable set'' is $x\in [1,5]$
instead of $x\in [1,1.2]$. Here the 
upper bound $5$ comes from
the constraint
$x\le 5$ that is in the parameter $M$ of the widening operator
$\nabla_{M}$. 
This
results in the graph 1 in Fig.~\ref{fig:iterseq}.

In the iteration sequence (Fig.~\ref{fig:iterseq}) the four 
polyhedra (in four different colors) gradually grow: in the graph 2 the water
level $x$ can be $10$~cm so in the graph 3 appears a green polyhedron
(meaning that a
signal is sent from the switch to the pump); after the graphs 3 and 9 we
\emph{delay} widening, a heuristic commonly employed in abstract
interpretation~\cite{Cousot1981}. In the end, in the graph 12 we have a
prefixed point (meaning that the polyhedra do not grow any further). There
we can see, from the range of $x$ spanned by the polyhedra, that the
water level never reaches beyond $0.6 \leq x \leq 12.2$. 


\subsection{Analysis by \emph{Nonstandard Abstract
  Interpretation}}\label{subsec:waterLevelDt}
In the above ``standard'' scenario, we approximated the 
dynamics of the water level by discretizing the continuous notion of
time ($\dt'=0.2$). While this made the usual abstract interpretation
workflow go around, 
there is a price to pay---the analysis result is not
\emph{precise}. Specifically, the reachable region thus over-approximated is
$0.6 \leq x \leq 12.2$, while the real reachable region is $1\leq x\leq
12$.\footnote{There are also examples in which 
discretization even leads to \emph{unsound} analysis results.}

\begin{wrapfigure}[13]{r}{0.5\hsize}
\vspace{-1em}
 \begingroup
 \fontsize{7pt}{8pt}\selectfont
\begin{verbatim}
(*Water-Level Monitor*)
l := 0; x := 1; p := 1; s := 0;
while true do {
   if p = 1 then x := x + dt 
      else x := x - 2 * dt;
   if (x <= 5 && p = 0) then s := 1 
      else {if (x >= 10 && p = 1) 
               then s := 1 
               else s := 0
      };
   if s = 1 then l := l + dt
      else skip;
   if s = 1 && l >= 2
      then {p := 1 - p; s := 0; l := 0}
      else skip
}
\end{verbatim}
\endgroup 
\vspace{-1.7em}
\caption{Water-level monitor in $\Whiledt$}
\label{fig:whileDtCodeCaseStudy}
\end{wrapfigure}
Obviously we can ``tighten up'' the analysis by making the value $\dt'$ smaller.
Even better, we can leave the expression $\dt'$ in
Fig.~\ref{fig:whileCodeCaseStudy} as a variable, and imagine the ``limit'' of
analysis results when the value of $\dt'$ tends to $0$. However here is
a question: what is that ``limit,'' in mathematically rigorous terms?
Taking $\dt'=0$ obviously does not work: do so in
Fig.~\ref{fig:whileCodeCaseStudy}
and we have no dynamics whatsoever. The value of $\dt'$ must be strictly positive.

Our contribution is an extension of abstract interpretation that answers
the last question. In our framework, the same (hybrid) dynamics of
the water-level monitor is modeled
by a program in Fig.~\ref{fig:whileDtCodeCaseStudy}. Here the expression
$\dt$
is a new constant that stands for a \emph{positive} and \emph{infinitesimal}
(i.e.\ infinitely small) value. Therefore the modeling is not an
approximation by discretization; it is an \emph{exact} modeling.

\begin{figure}[tb]
\begin{tabular}{lll}
\begin{minipage}[t]{.32\textwidth}
\includegraphics[width=\textwidth]{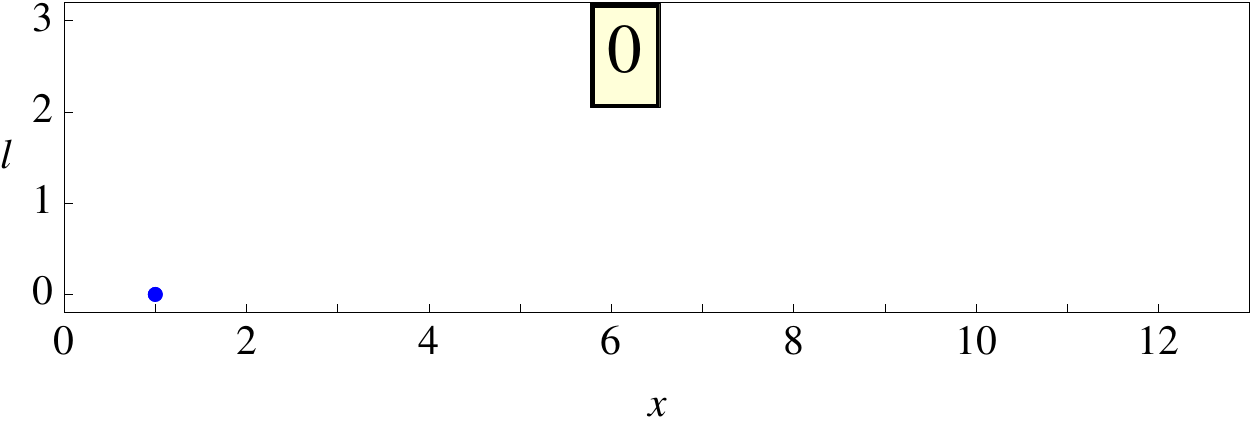}
\includegraphics[width=\textwidth]{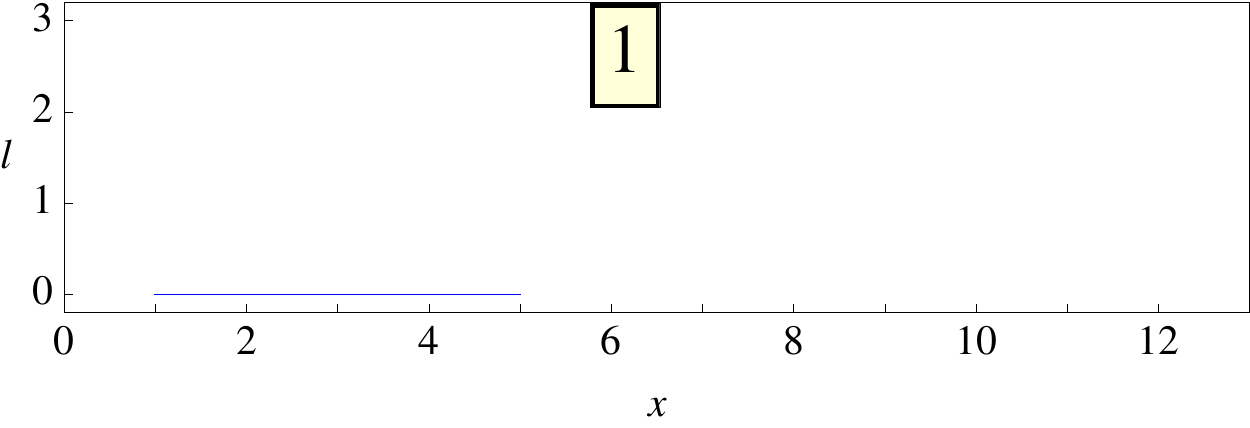}
\includegraphics[width=\textwidth]{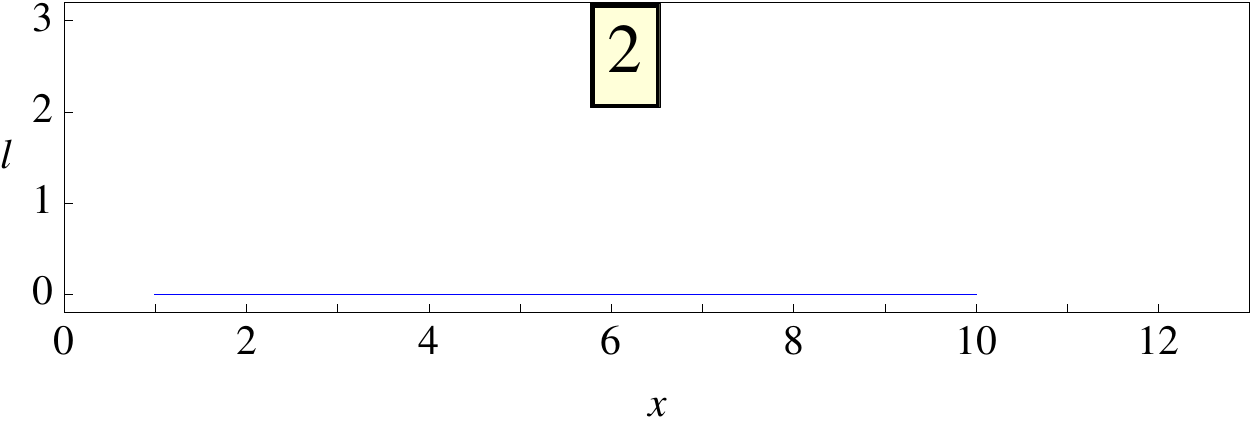}
\includegraphics[width=\textwidth]{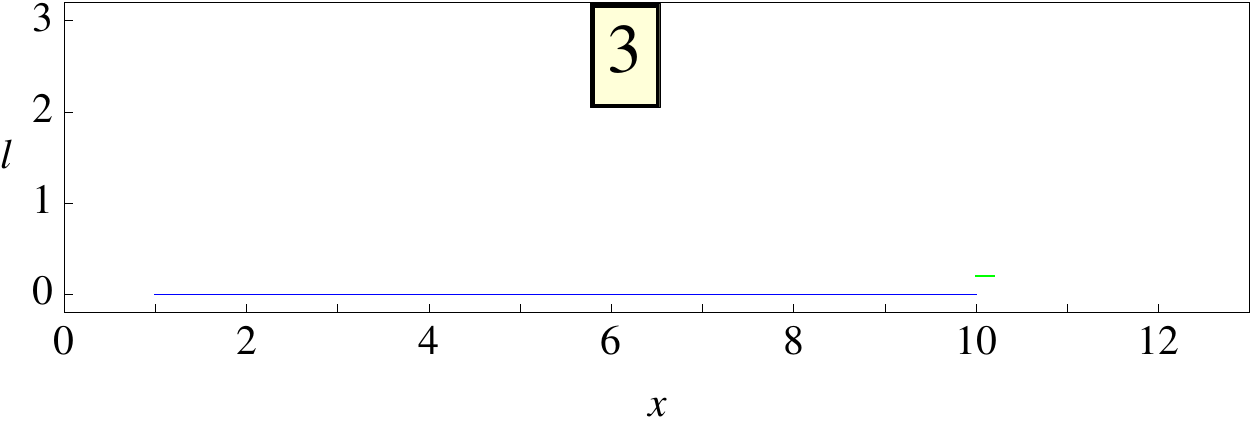}
\includegraphics[width=\textwidth]{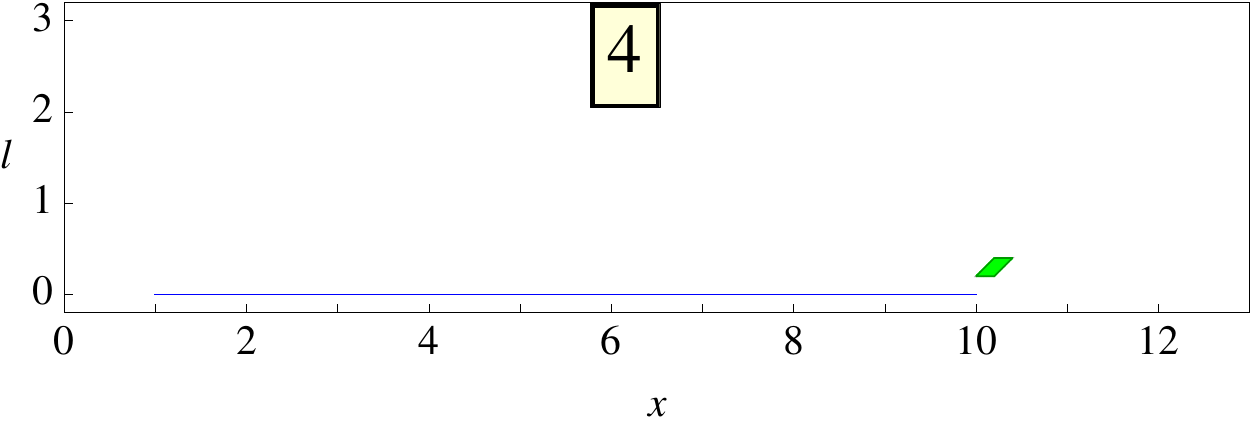}
\end{minipage}
\begin{minipage}[t]{.32\textwidth}
\includegraphics[width=\textwidth]{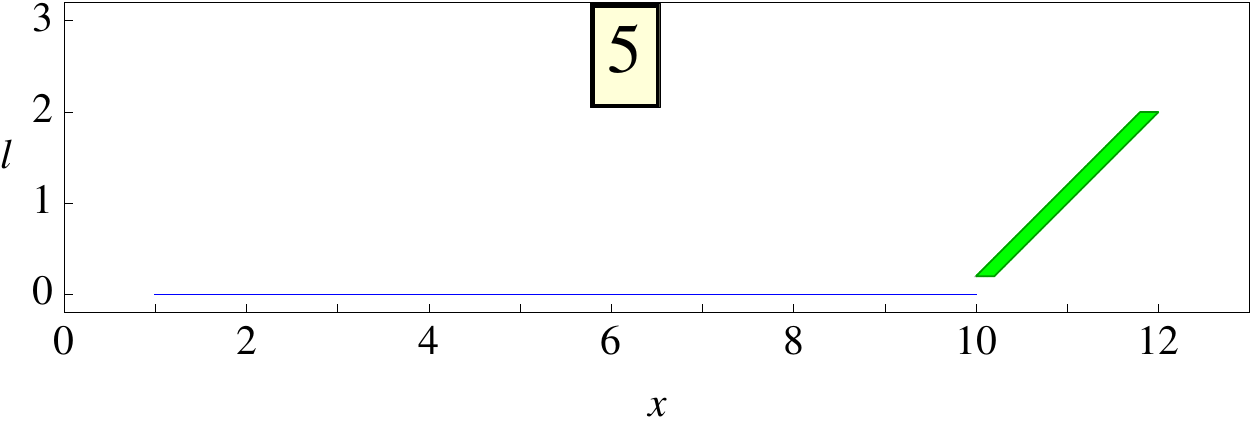}
\includegraphics[width=\textwidth]{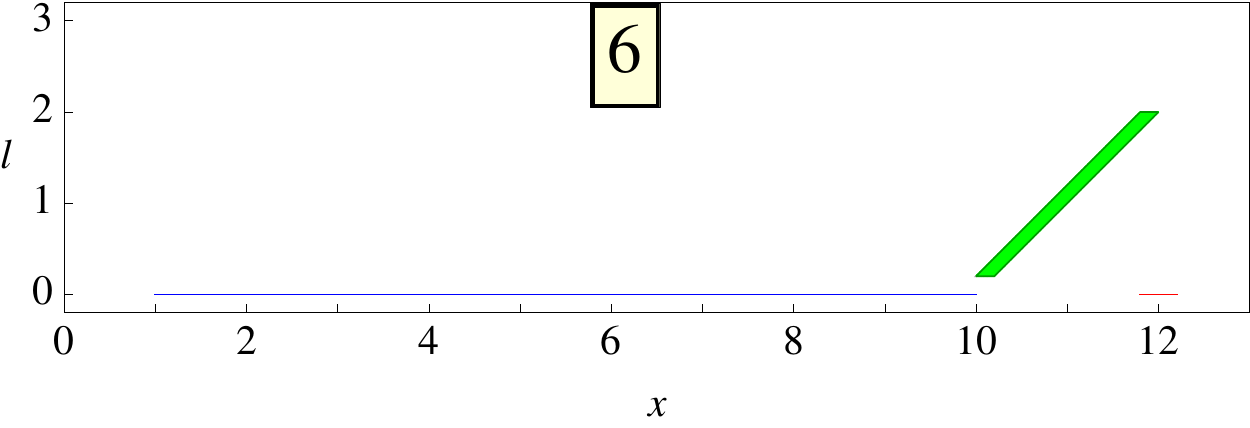}
\includegraphics[width=\textwidth]{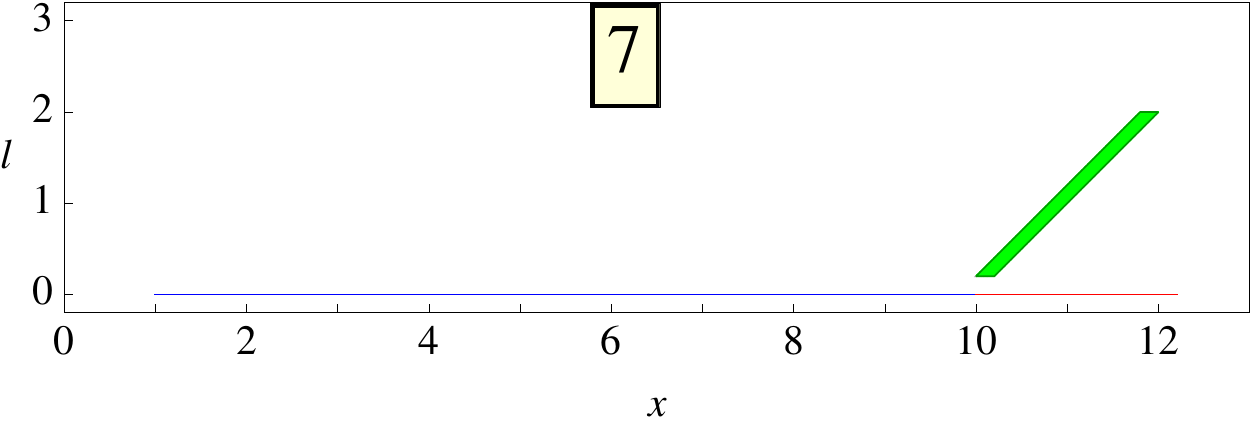}
\includegraphics[width=\textwidth]{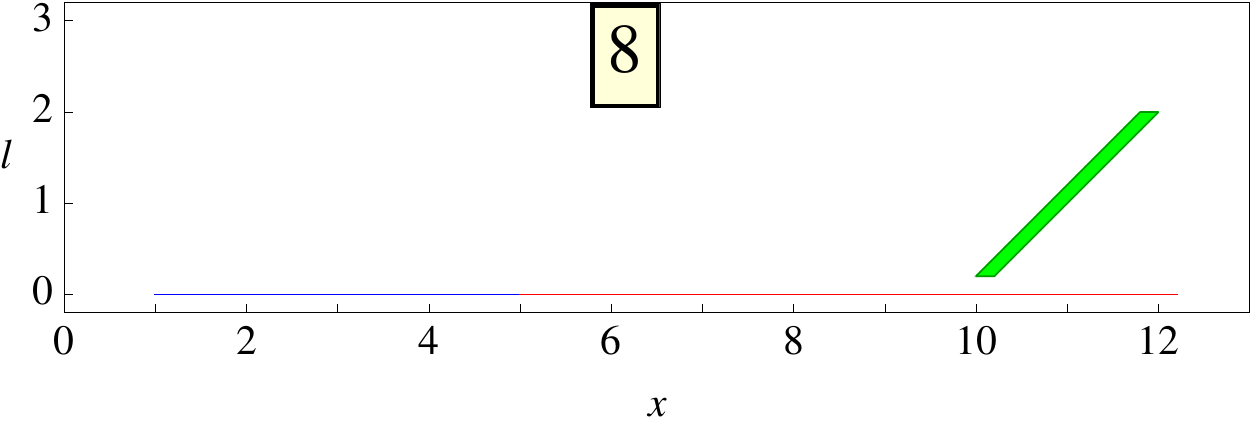}
\includegraphics[width=\textwidth]{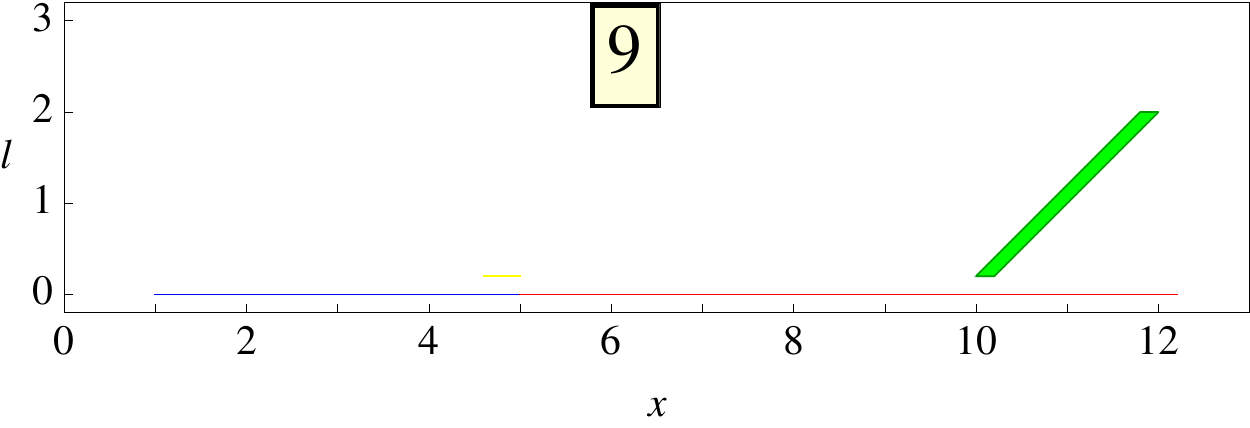}
\end{minipage}
\begin{minipage}[t]{.32\textwidth}
\includegraphics[width=\textwidth]{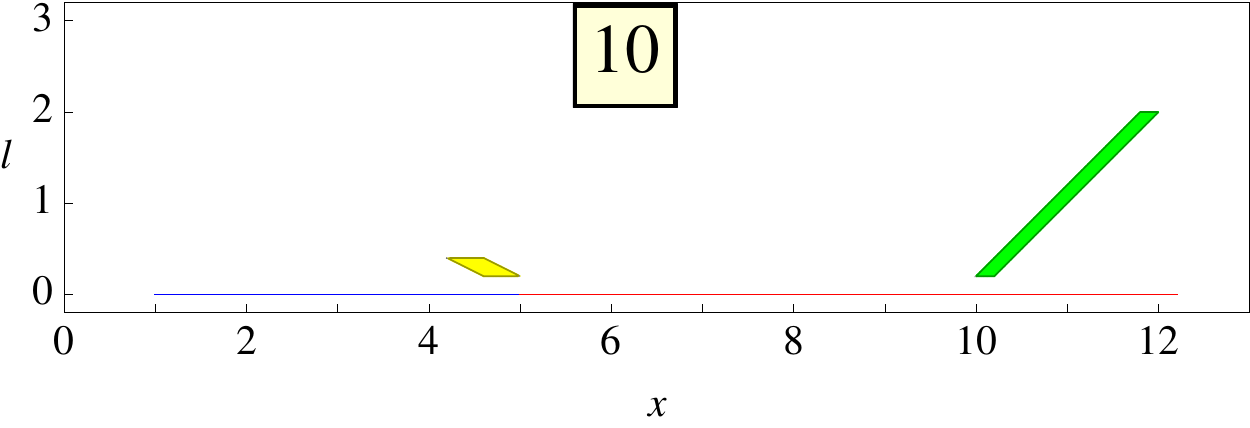}
\includegraphics[width=\textwidth]{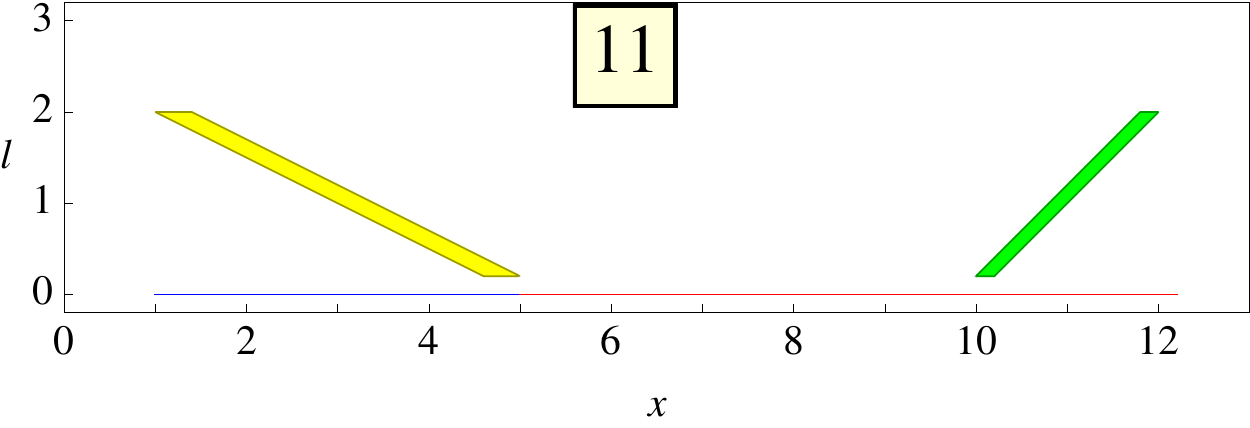}
\includegraphics[width=\textwidth]{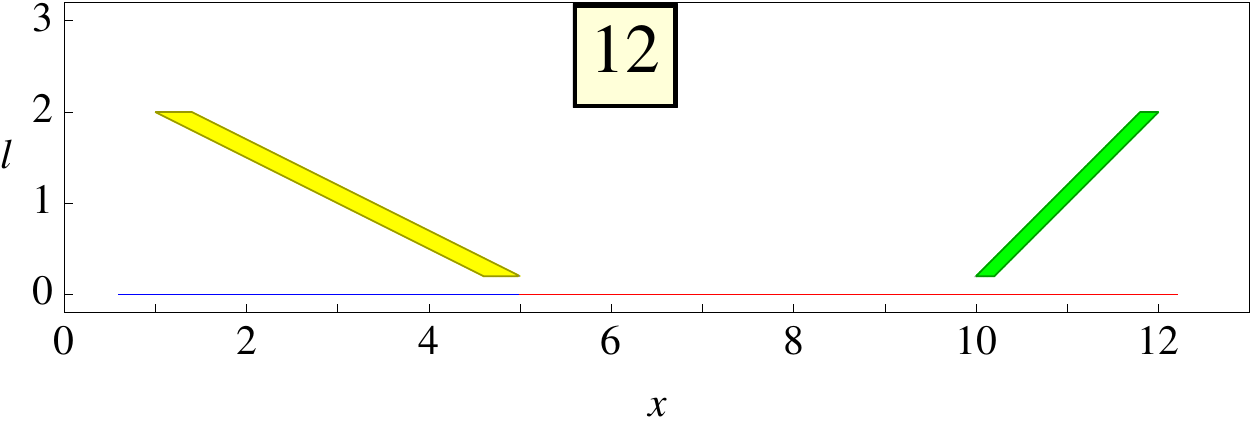}
\end{minipage}
\end{tabular}
 \caption{An iteration sequence for the water-level monitor example.\newline 
To save space, here we depict 
an element of $(\CP_{2})^{4}$---i.e.\ a quadruple of convex
polyhedra---on the same plane $\R^{2}$. The four convex polyhedra come in different colors:
those in blue, green, red and yellow correspond to the values
$(p,s)=(1,0), (1,1), (0,0)$ and $(0,1)$ of the Boolean variables,
respectively. 
} 
\label{fig:iterseq}
\end{figure}

It is important to notice that the program in
Fig.~\ref{fig:whileDtCodeCaseStudy} is the same as the one in
Fig.~\ref{fig:whileCodeCaseStudy}, except that now $\dt$ is some strange
constant,
while $\dt'$ in Fig.~\ref{fig:whileCodeCaseStudy} stood for a real
number (namely $0.2$). This difference, however, does not prevent us
from applying the \emph{static}, \emph{symbolic} and \emph{syntax-based} analysis by abstract
interpretation. We can follow exactly the same path as 
in~\S{}\ref{subsec:waterLevel0.2}---taking the abstract domain of convex
polyhedra, executing the program in Fig.~\ref{fig:whileDtCodeCaseStudy}
on it, applying the widening operator $\nabla_M$, and forming an
iteration sequence much like in Fig.~\ref{fig:iterseq}---and this leads
 to the analysis result $1-2\dt \leq x \leq 12+\dt$. Since $\dt$
is an infinitesimal number, the last result is practically as  good as
$1\le x \le 12$. We  have a prototype implementation that automates 
this analysis (\S{}\ref{sec:implementation}). 

What remains to be answered is the legitimacy of this extended abstract
interpretation framework. Is the outcome $1-2\dt \leq x \leq 12+\dt$
\emph{sound}, in the sense that it indeed over-approximates the true
reachable
set? Even before that, what do we mean by the ``true
reachable
set'' of the program in Fig.~\ref{fig:whileDtCodeCaseStudy}, 
with an exotic infinitesimal constant like $\dt$? Moreover,  are
iteration sequences via the widening operator $\nabla_M$ guaranteed to
terminate
within finitely many steps, as is the case in the standard
framework~\cite{Halbwachs1993, Halbwachs1997}?

The rest of the paper is mostly devoted to (answering positively to) the
last questions. In it we use Robinson's \emph{nonstandard analysis
(NSA)}~\cite{Robinson1966} and give  infinitesimal numbers---clearly
such do not exist in the set of (standard) real numbers---a status as first-class citizens. The program in
Fig.~\ref{fig:whileDtCodeCaseStudy} is in fact in the programming (or
rather \emph{modeling}) language $\Whiledt$ from~\cite{Suenaga2011,Hasuo2012};
and its semantics can be understood in the line of
Example~\ref{example:elapsedTime}. It turns out that the theory of
NSA---in particular its celebrated result of the \emph{transfer
principle}---allows us to ``transfer'' meta results from the standard
abstract interpretation to our extension. That is, what is true in the
world of standard reals (soundness, termination, etc.) is also true in 
that of \emph{hyperreals}.

%
%


\section{Preliminaries}\label{sec:preliminaries}
In~\S{}\ref{sec:NSAI}  we will present our \emph{soundness} and
\emph{termination} results as a ``metatheory'' that justifies the
workflow described in~\S{}\ref{subsec:waterLevelDt}; 
in this section we recall some preliminaries that are needed for those theoretical developments.
 First, the general theory of abstract interpretation is introduced in~\S{}\ref{subsec:abstinterp} and the specific domain of convex polyhedra is presented in~\S{}\ref{subsec:convexpoly}.
Next, some basic notions in nonstandard analysis are explained in~\S{}\ref{subsec:preliminariesNSA}.
Finally, in~\S{}\ref{subsec:whiledt}, the modeling language $\Whiledt$ from~\cite{Suenaga2011} and its (denotational) collecting semantics based on nonstandard analysis are presented.

\subsection{Abstract Interpretation}\label{subsec:abstinterp}
\emph{Abstract interpretation}~\cite{Cousot1978} is a well-established
technique in static analysis.  We make a brief review of its basic
theory; it is
mostly for the purpose of fixing notations.
The goal of abstract interpretation is to over-approximate a \emph{concrete semantics} defined on an \emph{concrete domain} by an \emph{abstract semantics} on an \emph{abstract domain}.
We assume that the concrete semantics is defined as a least fixed point on the concrete domain.
The following proposition guarantee the over-approximation of the least fixed point in the concrete domain by a prefixed point in the abstract domain.
In the proposition, the order $\sqsubseteq$ on the domain $L$ is extended to the order on $L\rightarrow L$ pointwisely.
And the \emph{least fixed point relative to $\basis$}, denoted by $\lfp_{\basis}F$, is the  
 least among the fixed points of $F$ above $\basis$; by the cpo
 structure of $L$ and the continuity of $F$, it is given by $\bigsqcup_{n\in\N}F^{n}\basis$. 
Note that we are using the concretization-based framework described in~\cite{Cousot1992a}.

\begin{myprop}\label{prop:concretization}
Let $(L, \sqsubseteq)$ be a cpo;
 $F:L\rightarrow L$ be a continuous function;
 and $\basis\in L$ be such that $\basis\sqsubseteq F(\basis)$.
Let $(\overline{L}, \overline{\sqsubseteq})$ be a preorder;
 $\gamma: \overline{L} \rightarrow L$ be a function (it is called \emph{concretization}) such that 
$\overline{a}\mathrel{\overline{\sqsubseteq}}\overline{b} \Rightarrow \gamma(\overline{a})\sqsubseteq \gamma(\overline{b})$ for all $\overline{a}, \overline{b} \in \overline{L}$; and
 $\overline{F}:\overline{L}\rightarrow\overline{L}$ be a monotone function such that $F\circ\gamma\sqsubseteq\gamma\circ\overline{F}$.
Assume further that  $\overline{x}\in\overline{L}$ is a prefixed point of  $\overline{F}$ 
 (i.e.\ $\overline{F}(\overline{x})\mathrel{\overline\sqsubseteq}\overline{x}$)
 such that $\basis\mathrel{\sqsubseteq}\gamma(\overline{x})$.

Then $\overline{x}$ over-approximates $\lfp_{\basis}F$, that is,
$\lfp_{\basis}F\sqsubseteq \gamma(\overline{x})$.
\myqed
\end{myprop}

In~\S{}\ref{subsec:waterLevel0.2} where we analyzed the discretized water-level monitor, the set $\Pow(\R^n)$ of subsets of memory states is used as a concrete domain $L$; and the \emph{domain of convex polyhedra} is used as an abstract domain $\overline{L}$. The interpretations $F$ and $\overline{F}$ on each domains are defined in a standard manner.
Towards the goal of obtaining $\overline{x}$ in Prop.~\ref{prop:concretization}, (i.e.\ finding a prefixed point in the abstract domain), the following notion of
\emph{widening} is used (often together with \emph{narrowing} that we will not
be using).
Note that in the following definition and proposition, the domain $(L, \sqsubseteq)$ is the abstract domain, corresponding to $(\overline{L}, \overline{\sqsubseteq})$ in Prop.~\ref{prop:concretization}.

\begin{mydef}[widening operator]\label{def:widen}
 Let $(L, \sqsubseteq)$ be a preorder.
 A function $\nabla: L \times L \rightarrow L$ is said to be a \emph{widening operator} if the following two conditions hold.
 \begin{itemize}
  \item (\emph{Covering}) For any $x, y \in L$, $x \sqsubseteq x \nabla y$ and $y \sqsubseteq x \nabla y$.
  \item (\emph{Termination}) For any ascending chain $\langle x_i
	\rangle \in L^\N$, the chain $\langle y_i \rangle \in
	L^\N$ defined by
$y_0 = x_0$
	 and
$y_{i+1} = y_{i} \nabla x_{i+1}$ for each $i \in \N$
	is ultimately stationary.
 \end{itemize}
\end{mydef}
\noindent  A widening operator on a fixed abstract domain $\overline{L}$
is not at all unique.  In this paper we will discuss three widening operators previously introduced for $\CP_{n}$.

 The use of widening is as in the following proposition: the covering
 condition  ensures that the outcome is a prefixed point; and the
 procedure terminates thanks to the
 termination condition.
\begin{myprop}[convergence of iteration sequences]\label{prop:widen}
 Let $(L, \sqsubseteq)$ be a preorder; $F: L\rightarrow L$ be a monotone function; $\basis \in L$ be such that $\basis \sqsubseteq F(\basis)$; $\nabla : L\times L \rightarrow L$ be a widening operator; and $\langle X_i \rangle_{i\in\N} \in L^{\N}$ be the infinite sequence defined by
 $$
\small
\begin{array}{c}
   X_0 = \basis\enspace; \qquad
   \text{and, for each $i\in\N$,}\quad
   X_{i+1} = 
   \begin{cases}
    X_{i} & (\text{if}\; F(X_i) \sqsubseteq X_{i})\\
    X_{i} \nabla F(X_{i}) & (otherwise)
   \end{cases}
\end{array}
$$
Then the sequence $\langle X_i \rangle_{i\in\N}$ is increasing and
 ultimately
 stationary; moreover its limit $\bigsqcup_{i\in\N}X_{n}$ is a prefixed
 point of $F$ such that $\basis\sqsubseteq\bigsqcup_{i\in\N}X_{n}$.
\myqed
\end{myprop}

\subsection{The Domain of Convex Polyhedra}\label{subsec:convexpoly}
The \emph{domain of convex polyhedra}, introduced in~\cite{Cousot1978},
is one of the most commonly used relational numerical abstract domains.
\begin{mydef}[domain of convex polyhedra $\CP_{n}$]\label{def:convexPoly}
An $n$-dimensional \emph{convex polyhedron} is the intersection of
 finitely many (closed) affine half-spaces.
We denote the set of convex polyhedra in $\R ^n$ by $\CP_n$.
Its preorder $\sqsubseteq$  is given by the inclusion order (actually it is a partial order).
The concretization function $\gamma_{\CP_n} : \CP_{n}\rightarrow\Pow(\R^n)$ is defined in an obvious manner.
\end{mydef}

\auxproof{
\begin{mydef}
A \emph{generator system} is a triple $(L, R, T)$ of finite sets of
 vectors.
Here 
$L$ 
is the set of \emph{lines}, $R$ is the set of \emph{rays} and $T$ is the
 set of \emph{points}.

A generator system $G=(\{\vec{l_1}, \cdots, \vec{l_l}\}, \{\vec{r_1},
 \cdots, \vec{r_r}\}, \{\vec{t_1}, \cdots, \vec{t_t}\})$ induces a
 convex polyhedron $\gen(G)$ in the following way, where
 $\R_{\geq 0}$ is the set of nonnegative reals.
$$
\gen(G)=\biggl\{\sum_{i=1}^l\lambda_i\vec{l_i} + \sum_{i=1}^r\rho_i\vec{r_i} + \sum_{i=1}^t\pi_i\vec{t_i} \,\biggr|\, \lambda_i \in \R, \rho_i \in \R_{\geq 0}, \pi_i \in \R_{\geq 0}, \sum_{i=1}^t \pi_i = 1 \biggr\}$$
\end{mydef}
}




We will be studying three widening operators on $\CP_{n}$. 
They are namely:
the \emph{standard widening operator} $\nabla_{S}$~\cite{Halbwachs1979};\footnote{The name ``standard'' is
 confusing with the distinction between \emph{standard} and
 \emph{nonstandard} entities in NSA. The  use of ``standard'' in the
 former sense is scarce in
 this paper.}
the \emph{widening operator $\nabla_M$ up to $M$}~\cite{Halbwachs1993, Halbwachs1997}; and
the \emph{precise widening operator} $\nabla_{N}$~\cite{Bagnara2005}.
We  briefly describe the former two; the definition of the last is
omitted for the lack of space. In the following definitions, the function $\con$ maps a set of linear constraints (called a \emph{constraint system}) to the convex polyhedron induced by the conjunction of those linear constraints.

\begin{mydef}[standard widening $\nabla_{S}$]\label{def:stdwiden}
 Let $P_{1}, P_{2}\in\CP_{n}$; and $C_{1}$ and $C_{2}$ be constraints
 system that induce $P_{1}$ and $P_{2}$, respectively.
 \emph{The standard widening operator} $\nabla_S : \CP_n \times \CP_n
 \rightarrow \CP_n$ is defined by

\scalebox{0.87}{\parbox{1.1\linewidth}{
 $$P_1 \nabla_S P_2 := \begin{cases}
			P_2 & \text{if\;}P_1 = \emptyset\\
			\con\Biggl(
			\begin{minipage}[c]{8cm}
			 $\{\varphi \in C_1 \mid
			 \text{$C_{2}$ implies $\varphi$, i.e.\
			 $\varphi$ is everywhere true in $P_{2}$}\}$\\
			 $\cup\bigl\{\psi \in C_2 \,\bigr|\, \exists \varphi \in C_1.\, P_1 = \con(C_1[\psi/\varphi])\bigr\}$
			\end{minipage}\Biggr)& \text{otherwise}.
		     \end{cases}$$
}}
\end{mydef}
\noindent Intuitively $P_1 \nabla_S P_2$ is represented by the set of those linear
constraints of $P_1$ which are satisfied by every point of $P_2$.

The following second widening operator $\nabla_{M}$ refines $\nabla_{S}$. This is
what we use in our implementation. Here $M$ is a parameter.
\begin{mydef}[widening up to $M$, $\nabla_{M}$]\label{def:widenupto}
 Let ${P}_1, {P}_2 \in \CP_n$, and $M$ be a (given) finite set of linear inequalities.
The \emph{widening operator up to $M$} is defined by
 $${P}_1 \nabla_M {P}_2 \;:=\; \bigl(\mathcal{P}_1 \nabla_S {P}_2\bigr)\; \cap\; \con\bigl(\{\varphi \in M \mid {P}_i\subseteq \con(\{\varphi\}){\rm \ for\ }i=1,2 \}\bigr)\enspace.$$
\end{mydef}
\noindent 
The parameter $M$   is usually taken to be the set of linear
inequalities that occur in the  program under analysis.


\subsection{Nonstandard Analysis}\label{subsec:preliminariesNSA}
Here we list a minimal set of necessary definitions and results in
nonstandard analysis (NSA)~\cite{Robinson1966}.
Some further details 
can be found in Appendix~\ref{appendix:NSAPrimer}; 
fully-fledged and accessible expositions of NSA are 
found e.g.\ in~\cite{Hurd1985, Goldblatt1998}.

The following notions will play important roles.
\begin{itemize}
 \item 
 \emph{Hyperreals} that extends reals by infinitesimals, infinites,
       etc.;
 \item 
 The \emph{transfer principle}, a celebrated result in NSA that states
 that
 reals and hyperreals share ``the same properties''; 
 \item 
 The first-order language $\LX$ that specifies formulas in which syntax, precisely,
 are preserved by the transfer principle; and finally
 \item 
 The semantical construct of \emph{superstructure} for
 interpreting $\LX$-formulas.
\end{itemize}
What is of paramount importance is the transfer principle; in order to 
formulate it in a mathematically rigorous manner, 
the two last items (the language $\LX$ on the syntactic side, and
superstructures on the semantical side) are used.
The first-order language $\LX$ is essentially that of set theory and has
two predicates $=$ and $\in$. The \emph{superstructure} $V(\baseSet)$ is
then a semantical ``universe'' for such formulas, constructed from the
base set $X$: concretely $V(\baseSet)$ is the union of $X$, $\pow(X)$,
$\pow(X\cup \pow(X))$, and so on. Finally, when we take $X=\R$ then the
set $\hyper{X}=\hyper{\R}$ is that of \emph{hyperreals}; and the
transfer principle claims that $A$ holds for reals if and only if
$\hyper{A}$---a formula essentially the same as $A$---holds for
hyperreals.
 Its precise
statement is:
\begin{mylem}[the transfer principle]\label{lemma:transferPrinciplePreview}
 For any closed formula $A$ in $\LX$, the following are equivalent. 
\begin{itemize}
 \item The formula $A$ is valid in the superstructure $V(\baseSet)$.
 \item The \emph{*-transform} $\hyper{A}$ of $A$---this is a formula in the
       language $\LsX$---is valid in the superstructure $V(\hyper{\baseSet})$.
\end{itemize}
\end{mylem}

The transfer principle guarantees that we can employ the same abstract
 interpretation
framework, for reals and hyperreals alike---\emph{literally} the same, in the sense that we express the framework
 in the
 language $\LR$. Concretely, various
constructions
and meta results (such as soundness and termination) in abstract
interpretation will be expressed as $\LR$-formulas, and since they are valid
 in $V(\R)$, they are valid in the ``nonstandard universe''
 $V(\hyper{\R})$ too, by the transfer principle.

\paragraph{Hyperreals}
We fix an \emph{index set} $I=\N$, and an \emph{ultrafilter} $\filt\subseteq\pow(I)$ that extends
 the cofinite filter $\filt_{\mathrm{c}}:=\{S\subseteq I\mid
 I\setminus S \text{ is finite}\}$. Its properties to be noted: 1)
 for any $S\subseteq I$, exactly one of $S$ and $I\setminus S$
 belongs to $\filt$; 2) if $S$ is \emph{cofinite} (i.e.\ $I\setminus
 S$ is finite), then $S$ belongs to $\filt$.
 
\begin{mydef}[hyperreal
 $r\in\hyper{\R}$]\label{definition:hypernumber}
We define the set
 $\hyper{\R}$ of \emph{hyperreal numbers} (or \emph{hyperreals})
by
 \begin{math}
  \hyper{\R}:=\R^{I}/{\sim_{\filt}}
 \end{math}. It is therefore the set of infinite sequences on $\R$ modulo the
 following equivalence $\sim_{\filt}$: we have
 $(a_{0},a_{1},\dotsc)\sim_{\filt} (a'_{0},a'_{1},\dotsc)$ if
 \begin{equation}\label{equation:defOfSimFilt}
 \{i\in I\mid a_{i}=a'_{i}\}\in\filt\enspace,
 \quad\text{for which we say ``$d_{i}=d'_{i}$ for almost every $i$.''}
 \end{equation}
A \emph{hypernatural} $n\in\hyper{\N}$ is defined similarly.
 \end{mydef}
\noindent
It follows that:
 two sequences $(a_{i})_{i}$ and $(a'_{i})_{i}$ that
coincide except for finitely many indices $i$ represent the
same hyperreal.  The predicates besides $=$  (such as $<$) are defined in the same
way.
A notable consequence is the existence of infinite numbers in the set of hyperreals and hypernaturals:
$\omega:=[(1, 2, 3, \dotsc)]$ is a positive infinite since it is larger than any positive real $r = [(r, r, \dotsc)]$ ($i > r$ for almost every $i\in\N$).
In addition, the set of hyperreals includes infinitesimal numbers: a hyperreal
 $\omega^{-1}:=[\,(1,\frac{1}{2},\frac{1}{3},\dotsc)\,]$ is positive
 ($0<\omega^{-1}$) but is smaller than any (standard) positive real
 $r$.



\paragraph{Superstructure}

A \emph{superstructure} is a ``universe,'' constructed step by
step
 from
a certain base set $\baseSet$ (whose typical examples are
$\R$ and $\hyper{\R}$). We assume
 $\N\subseteq \baseSet$.
\begin{mydef}[superstructure]\label{definition:superstructure}
A \emph{superstructure} $V(\baseSet)$ over $\baseSet$ is
 defined by $  V(\baseSet):=\bigcup_{n\in\N}V_{n}(\baseSet)$, where
$V_{0}(\baseSet):=\baseSet$ and
$V_{n+1}(\baseSet) := V_{n}(\baseSet) \cup \pow (V_{n}(\baseSet))$.
 \end{mydef}

The superstructure $V(X)$ might seem to be a closure of $X$ only under powersets, but it accommodates many set-forming operations.
For example, ordered pairs $\rtuple{a,b}$ and tuples $\rtuple{a_{1},\dotsc,a_{m}}$ are 
defined in $V(\baseSet)$ as is usually done in set theory, e.g.\
$\rtuple{a,b}:=\{\{a\},\{a,b\}\}$.
The function space $a\to b$ is
thought of as a collection of special binary relations (i.e.\ $a\to
b\subseteq \pow (a\times b))$,  hence is in $V(\baseSet)$.

\paragraph{The First-Order Language $\LX$}
We use the following 
first-order language $\LX$, defined for each choice of the base set 
$X$ like $\R$ and $\hyper{\R}$.
\begin{mydef}[the language $\LX$]\label{definition:languageLX}
 \emph{Terms} in $\LX$ consist of: variables $x,y,x_{1},x_{2},\dotsc$;
 and a constant $a$ for each entity $a\in V(\baseSet)$. 

\emph{Formulas} in $\LX$ are constructed as follows.
\begin{itemize}
 \item The predicate symbols are $=$ and $\in$; both are binary. The
       \emph{atomic formulas} are of the form $s=t$ or $s\in t$ (where $s$
       and $t$ are terms).
 \item We allow  Boolean combinations of formulas. We use the
       symbols
       $\land, \lor,\lnot$ and $\Rightarrow$.
 \item Given a formula $A$, a variable $x$ and a term $s$, the expressions
    $\forall x\in s.\, A$ and     $\exists x\in s.\, A$
       are formulas. 
\end{itemize}
\end{mydef}
\noindent Note that
 quantifiers always come with a bound $s$.
 The language $\LX$ depends on the choice of $\baseSet$ (it
  determines the set of constants). 
We shall also  use
 the following syntax sugars in $\LX$, as is common in set theory and NSA.  
 \begin{displaymath}\footnotesize
  \begin{array}{ll}
   \rtuple{s,t} & \text{pair}
  \qquad  \qquad  \qquad
   \rtuple{s_{1},\dotsc,s_{m}} \quad \text{tuple}
  \qquad  \qquad  \qquad
 s\times t \quad \text{direct product}
  \\
   s\subseteq t &\text{inclusion, short for $\forall x\in s. \,x\in t$}
  \\
   s(t) &\text{function application; short for $x$ such that
    $\rtuple{t,x}\in s$}
  \\
   s\co t &\text{function composition, $(s\co t)(x)=s(t(x))$}
  \\
   s\le t&\text{inequality in $\N$; short for  $\rtuple{s,t}\in{\le}$
    where ${\le}\subseteq\N^{2}$}
  \end{array}
 \end{displaymath}


\begin{mydef}[semantics of $\LX$]\label{definition:semanticsOfLX}
 We interpret $\LX$ in the superstructure $V(\baseSet)$ in the obvious way.
Let $A$ be a  closed formula;  we say $A$ is \emph{valid}
if $A$ is true in $V(\baseSet)$.
\end{mydef}

\paragraph{The $*$-Transform and the Transfer Principle}
 As we mentioned the transfer principle says that a closed formula $A$
 in the language $\LX$ is valid in $V(X)$ if and only if $\hyper{A}$ in
 $\LsX$ is valid in $V(\hyper{X})$.
  We shall describe how we syntactically transform
$A$ in $\LX$ into
 $\hyper{A}$ in $\LsX$. 

 For that purpose, in particular in translating constants in $\LX$ (for entities in
 $V(X)$) to $\LsX$,  we will need the following \emph{semantical} translation. 
 The so-called \emph{ultrapower construction} yields a canonical map
 \begin{equation}\label{equation:*TransferMap}\footnotesize
\begin{array}{c}
   \hyper{(\place)}
  \;:\; 
  V(\baseSet) \longrightarrow  V(\hyper{\baseSet})\enspace,
  \qquad
  a \longmapsto \hyper{a}\enspace
\end{array} 
\end{equation}
that is called the \emph{*-transform}. 
It is a map from the universe $V(\baseSet)$ of standard entities
to $V(\hyper{\baseSet})$ of nonstandard entities.  The details of its
 construction are in Appendix~\ref{appendix:NSAPrimer} or in ~\cite{Hurd1985}. 
 

 The above map $\hyper{(\place)}\colon V(X)\to V(\hyper{X})$ becomes a \emph{monomorphism}, a notion in
 NSA.  Most notably it will satisfy
 the \emph{transfer principle} (Lem.~\ref{lemma:transferPrinciple}).

\begin{mydef}[*-transform of formulas]\label{definition:starTransformOfFormulas}
 Let $A$ be a formula in $\LX$. The \emph{*-transform} of $A$, denoted
 by $\hyper{A}$, is 
 a formula in $\LsX$ obtained by replacing each constant $a$ occurring
 in $A$ with the constant $\hyper{a}$ that designates the element $\hyper{a}\in
 V(\hyper{\baseSet})$.
\end{mydef}


\begin{mylem}[the transfer principle]\label{lemma:transferPrinciple}
 For any closed formula $A$ in $\LX$, $A$ is valid (in $V(\baseSet)$) if and
 only if $\hyper{A}$ is valid (in $V(\hyper{\baseSet})$).  \myqed
\end{mylem}

We can prove, for instance, the following proposition using the transfer
principle (the proof is in Appendix~\ref{appendix:omittedProofs}). 
This proposition has a practical implication: our implementation relies
on it 
in simplifying formulas including the infinitesimal constant $\dt$.

\begin{myprop}\label{prop:dtToQE}
 Let $A$ be an $\LR$-formula with a unique free variable $x$; to
 emphasize it we write $A(x)$ for $A$. 
Then the validity of the formula
$$\exists r \in \R. \left(0<r \wedge \forall x \in \R. \left(0<x<r \Rightarrow A \left(x\right)\right)\right)$$
(in $V(\R)$) implies the validity of $\hyper A (\dt)$ in $V(\hyper{\R})$.
\myqed
\end{myprop}

\subsection{The Modeling Language $\Whiledt$}\label{subsec:whiledt}

$\Whiledt$, a modeling language for hybrid systems based on NSA, is introduced in \cite{Suenaga2011}.
It is an augmentation of a usual imperative language (such as ${\mathbf{IMP}}$ in~\cite{Winskel1993}) with a constant $\dt$ that expresses an infinitesimal number.

\begin{mydef}
Let $\Var$ be the set of variables.
The syntax of $\Whiledt$ is as follows:
\begin{eqnarray*}\footnotesize
\begin{array}[tb]{rl}
\AExp \ni a ::=& x \mid r \mid a_1 \aop a_2 \mid \dt\\
&\text{\; where\; } x \in \Var, r \in \R \text{\; and\; } \aop \in \{+, -, \cdot, / \}\\
\BExp \ni b ::=& \true \mid \false \mid b_1 \wedge b_2 \mid \neg b \mid a_1 < a_2  \\
\Cmd \ni c ::=& \SKIP \mid x:=a \mid c_1;c_2 \mid \IF\; b\; \THEN\; c_1\; \ELSE\; c_2 \mid \WHILE\; b\; \DO\; c. 
\end{array}
\end{eqnarray*}
An expression $a \in \AExp$ is an \emph{arithmetic expression}, $b \in \BExp$ is a \emph{Boolean expression} and $c \in \Cmd$ is a \emph{command}.
\end{mydef}

\begin{wrapfigure}[9]{r}{0.46\hsize}
\vspace{-1em}
 \begingroup
 \fontsize{7pt}{8pt}\selectfont
  \begin{verbatim}
(*Thermostat*)
x := 22; p := 0;
while true do {
   if p = 0 then x := x - 3 * x * dt 
      else x := x + 3 * (30 - x) * dt;
   if x >= 22 then p := 0
      else {if x <= 18 then p := 1
               else skip
      }
}
\end{verbatim}
\endgroup 
 \vspace{-1.7em}
\caption{Thermostat in $\Whiledt$}
\label{fig:thermostat} 
\end{wrapfigure}
As we explained in~\S{}\ref{sec:introduction}, the infinitesimal constant $\dt$ enables us to model not only discrete dynamics but also continuous dynamics without explicit ODEs.
For example, the water-level monitor is modeled as a $\Whiledt$ program shown in Fig.~\ref{fig:whileDtCodeCaseStudy}.
As another example, the thermostat can be modeled as the program on the right.
One can see that the continuous dynamics modeled in this example is beyond piecewise-linear.
Even dynamics defined by nonlinear ODEs can be modeled in $\Whiledt$ in
the same manner. To go further to accommodate an arbitrary hybrid automaton we must
properly deal with \emph{nondeterminism}, a feature currently lacking in
$\Whiledt$. Although we expect that to be not hard, precise comparision
between $\Whiledt$ and hybrid automata in expressivity is future work.

In the usual, standard abstract interpretation (without $\dt$), a
command $c$ is assigned its \emph{collecting semantics}
$\Pow(\Var\rightarrow\R) \rightarrow \Pow(\Var\rightarrow\R)$ (see e.g.~\cite{Cousot1977}). This is semantics by reachable sets of memory states,  as the concrete semantics.
 Presence of $\dt$ in the syntax of $\Whiledt$  calls for an
infinitesimal number in the picture.
The first thing to try would be to replace $\R$ with $\hyper\R$, and
let  $\Whiledt$ commands interpreted as functions of  the type
$\Pow(\Var\rightarrow\hyper{\R}) \rightarrow
\Pow(\Var\rightarrow\hyper{\R})$. This however is not suited for the
purpose of interpreting recursion in presence of $\dt$.\footnote{If we interpret commands
as functions  $
\Pow(\Var\rightarrow\hyper{\R})
\to
\Pow(\Var\rightarrow\hyper{\R})$, the interpretation
$\sem{\WHILE\; x<10 \;\DO\; x:= x+\dt}\{(x\mapsto 0)\}$ by a least fixed point
will be $\{x\mapsto r \mid \exists n \in \N.\; r = n*\dt\}$, not
$\{x\mapsto r \mid \exists n \in \hyper\N.\; r = n*\dt \wedge
r\leq10\}$ as we expect. The problem is that
\emph{internality}---an ``well-behavedness'' notion in
NSA---is not preserved in such a modeling.} We rely instead on our theory of
\emph{hyperdomains} that is used in~\cite{Suenaga2013} and
 described in Appendix~\ref{appendix:domainTheoryTransferred}
; see the
interpretation of while loops in Table~\ref{table:densem}. This calls
for the interpretation of commands to be of the type $\hyper{\bigl(\,
\Pow(\Var\rightarrow\R) \rightarrow \Pow(\Var\rightarrow\R)
\,\bigr)}$,  a subset of 
$\hyper{
\Pow(\Var\rightarrow\R)} \rightarrow \hyper{\Pow(\Var\rightarrow\R)
}$. The last type will be used in the following definition.

\begin{mydef}
\label{def:whiledtsem}
 \emph{Collecting semantics} for $\Whiledt$,  
in Table~\ref{table:densem}, has the following
 types where $\B$ is $\{\ttrue, \ffalse\}$: 
$\sem{a}\colon \hyper(\Var\rightarrow\R) \rightarrow \hyper\R $ for
 $a\in\AExp$;
$\sem{b}\colon \hyper(\Var\rightarrow\R) \rightarrow \B$ for
 $b\in\BExp$; and
$\sem{c}\colon \hyper\Pow(\Var\rightarrow\R) \rightarrow
 \hyper\Pow(\Var\rightarrow\R)$ for $c\in\Cmd$.
\end{mydef}

In~\cite{Suenaga2011} and in~\S{}\ref{sec:introduction}, the
 semantics of a while loop is defined using the idea of
sectionwise execution, instead of as a least fixed point. This is not
suited for  employing abstract
interpretation---the latter is after all for computing least fixed points.
The collecting semantics in Def.~\ref{def:whiledtsem}
(Table~\ref{table:densem}) does use least fixed
points; it is based on the alternative $\Whiledt$ semantics 
introduced in~\cite{Kido2013} (it will also appear in the forthcoming full version
of~\cite{Suenaga2011, Hasuo2012}). The equivalence of the two semantics 
is established in~\cite{Kido2013}.



\begin{table}[tbp]
 \centering
\scalebox{0.8}{\parbox{1.1\linewidth}{%
 \begin{align*}
 & \sem{x}\hsigma := \hsigma(x) \text{\;for each\;}x\in\Var  &&\sem{\true}\hsigma := {\ttrue} \\
 & \sem{r}\hsigma := r \text{\;for each\;}r \in\R  &&\sem{\false}\hsigma := {\ffalse} \\
 & \sem{a_1 \aop a_2}\hsigma := \sem{a_1} \aop \sem{a_2} &&\sem{b_1 \wedge b_2}\hsigma := \sem{b_1} \wedge \sem{b_2} \\
 & \sem{\dt}\hsigma := \textstyle[(1, \frac{1}{2}, \frac{1}{3}, \cdots)]  &&\sem{\neg b}\hsigma := \neg(\sem{b}\hsigma)
 \end{align*}

 \vspace{-2.5em}
 \begin{align*}
 &\sem{\SKIP}\hS := \hS \\
 &\sem{x := a}\hS := \{\hsigma[\sem{a}\hsigma/x] \mid \hsigma \in \hS\} \\
 &\sem{c_1; c_2}\hS := \sem{c_2}(\sem{c_1}\hS) \\
 &\sem{\IF\; b \;\THEN\; c_1 \;\ELSE\; c_2}\hS :=
 \begin{minipage}[c]{8cm}
  $\{\sem{c_1}\hsigma \mid \hsigma\in \hS,\; \sem{b}\hsigma = \ttrue \} \\ \cup \{\sem{c_2}\hsigma \mid \hsigma\in \hS,\; \sem{b}\hsigma = \ffalse \}$
 \end{minipage}\\
 &\sem{\WHILE\; b \;\DO\; c} := \lfp\bigl(\hyper\Phi\left(\sem{b}\right)\left(\sem{c}\right)\bigr) \\
 &\qquad\text{where\;}\Phi: \left(\St \rightarrow \B\cup\{\bot\}\right) \rightarrow \bigl(\Pow\left(\Var\rightarrow\R\right)\rightarrow \Pow\left(\Var\rightarrow\R\right)\bigr) \rightarrow \\
 &\qquad\qquad\quad \Bigl(\bigl(\Pow\left(\Var\rightarrow\R\right)\rightarrow\Pow\left(\Var\rightarrow\R\right)\bigr) \rightarrow \bigl(\Pow\left(\Var\rightarrow\R\right)\rightarrow\Pow\left(\Var\rightarrow\R\right)\bigr)\Bigr)\\
 &\qquad\text{\;is\;defined\;by\;}
 \Phi(f)(g) = \lambda\psi.\;\lambda S.\;
 S\cup\psi\{(g(\sigma))\mid \sigma\in S,\; f(\sigma) = \ttrue\} \cup \{\sigma\mid\sigma\in S, \; f(\sigma) = \ffalse\}.
 \end{align*}
}}
\caption{$\Whiledt$ collecting semantics}
\label{table:densem}
\end{table}

In the rest of the paper we restrict the set of variables $\Var$ to be
finite. This as\-sump\-tion---a realistic one when we focus on the program
to be analyzed---makes our NSA framework much simpler. 
Therefore
$\Pow(\Var\rightarrow \R)$ and $\hyper\Pow(\Var\rightarrow\R)$ are equal to $\Pow(\R^n)$ and $\hyper\Pow(\R^n)$ for some $n\in\N$ respectively; we prefer the latter notations in what follows.



\section{Abstract Interpretation Augmented with Infinitesimals}
\label{sec:NSAI} 
In the current section are our main theoretical contributions---a
metatheory of \emph{nonstandard abstract interpretation} that justifies 
the workflow in~\S{}\ref{subsec:waterLevelDt}. 

(Standard) abstract interpretation infrastructure such as Prop.~\ref{prop:concretization} and Prop.~\ref{prop:widen} is not applicable to $\Whiledt$ programs.
since $\hyper\Pow(\R^n)$ is not a cpo.\footnote{One can see that the ascending chain defined by $X_n := \{k*\dt \mid 0\leq k \leq n\}$ does not have the supremum in $\hyper\Pow(\R^n)$ since $\{k*\dt \mid k\in\N\}$ is not \emph{internal}  (see Appendix~\ref{appendix:NSAPrimer})
.}
Thus, building on the theoretical foundations in the above, we now extend the abstract interpretation framework for the
analysis of $\Whiledt$ programs (and the hybrid systems modeled
thereby).  We introduce an \emph{abstract hyperdomain} over $\hyper{\R}$  as the transfer of
 the 
(standard, over $\R$) domain of convex polyhedra. We then interpret $\Whiledt$
programs in them, and transfer the three widening operators mentioned in~\S{}\ref{subsec:abstinterp} to the
nonstandard setting. We classify them into \emph{uniform} ones---for
which termination is guaranteed even in the nonstandard setting---and
non-uniform ones.
The main theorems are Thm.~\ref{thm:hyperconcretization} and Thm.~\ref{thm:newunifwidenwithinfinitesimal}, for soundness (in place of Prop.~\ref{prop:concretization}) and termination (in place of Prop.~\ref{prop:widen}) respectively.


\subsection{The Domain of Convex Polyhedra over Hyperreals}
\label{subsec:abstractDomainOverHyperreals}

We extend   convex polyhedra to the current nonstandard setting.
\begin{mydef}[convex polyhedra over $\hyper{\R}$]\label{def:hyperConvexPoly}
 A \emph{convex polyhedron} on $(\hyper\R)^n$ is an intersection of
 finite number of affine half-spaces  on $(\hyper\R)^n$, that is, the
 set of points $\mathbf{x}\in(\hyper\R)^n$ that satisfy a certain
 finite set of linear inequalities.
The set of all convex polyhedra on $(\hyper\R)^n$ is denoted by
 $\CP_n^{\hyper{\R}}$.
 
\end{mydef}


\begin{myprop}\label{prop:hyperconvexpoly}
The set $\CP_n^{\hyper{\R}}$ of all  convex  polyhedra over $(\hyper{\R})^{n}$
is a (proper) subset of
$\hyper\CP_n$, the $*$-transform of the (standard) domain of convex polyhedra over $\R^n$.
\myqed
\end{myprop}
What lies in the difference between the two sets
$\CP_n^{\hyper{\R}}\subsetneq\hyper\CP_n$ is, for example, a disk as a
subset of $\R^{2}$ (hence of $\hyper{\R}^{2}$). In $\hyper\CP_2$ one can use a constraint system whose number of linear constraints is a hypernatural number $m\in\hyper{\N}$; using e.g.\
$m=\omega=[(0,1,2,\dotsc)]$
allows us to approximate a disk with progressive precision.

In the following development of nonstandard abstract interpretation,  we
will use $\hyper\CP_n$ as an abstract domain since it allows transfer
of properties 
of $\CP_n$.
 We note, however, that 
our over-approximation of
the interpretation $\sem{c}$ of a loop-free $\Whiledt$ program $c$ 
is always given in $\CP_n^{\hyper{\R}}$, i.e.\ with finitely many linear inequalities.


\subsection{Theory of Nonstandard Abstract Interpretation}
\label{subsec:theoryOfHyperAbstractInterp}

Our goal is to over-approximate the collecting semantics for $\Whiledt$ programs (Table~\ref{table:densem}) on convex polyhedra over $\hyper\R$.
As we mentioned at the beginning of this section, however, abstract interpretation infrastructure cannot be applied since $\hyper\Pow(\R^n)$ is not a cpo.
Fortunately it turns out that we can rely on the \emph{$*$-transform}
(\S{}\ref{subsec:preliminariesNSA}) of
the theory in~\S{}\ref{subsec:abstinterp}, where it suffices to impose the cpo structure only on $\Pow(\R)$ and the \emph{$*$-continuity}---instead of the (standard)
continuity---on
the function
 $\sem{c}$.
 This theoretical framework of \emph{nonstandard abstract
 interpretation}, which we shall describe here, is an extension of the
 \emph{transferred domain theory} studied
 in~\cite{BeauxisM11,Suenaga2013}. Part of the latter is found also in Appendix~\ref{appendix:domainTheoryTransferred}.

\begin{mythm}\label{thm:hyperconcretization}
Let $(L, \sqsubseteq)$ be a cpo;
 $F:\hyper{L}\rightarrow \hyper{L}$ be a *-continuous function;
 and $\basis\in\hyper{L}$ be such that $\basis\mathrel{\hyper\sqsubseteq} F(\basis)$.
Let $(\overline{L}, \overline{\sqsubseteq})$ be a preorder;
 $\gamma: \overline{L} \rightarrow L$ be a function such that $\overline{a}\mathrel{\overline{\sqsubseteq}}\overline{b} \Rightarrow \gamma(\overline{a})\sqsubseteq \gamma(\overline{b})$ for all $\overline{a}, \overline{b}\in\overline{L}$; and
 $\overline{F}:\hyper{\overline{L}}\rightarrow\hyper{\overline{L}}$ be a *-continuous function that is monotone with respect to $\hyper{\overline\sqsubseteq}$ and satisfies $F\circ\hyper\gamma \mathrel{\hyper\sqsubseteq} \hyper\gamma\circ\overline{F}$.
 Note that $(\hyper{\overline{L}}, \hyper{\overline\sqsubseteq})$ is also a preorder.
Assume further that  $\overline{x}\in\hyper{\overline{L}}$ is a prefixed point of  $\overline{F}$ 
 (i.e.\ $\overline{F}(\overline{x})\mathrel{\hyper{\overline\sqsubseteq}}\overline{x}$)
 such that $\basis\mathrel{\hyper{\sqsubseteq}}\hyper{\gamma}(\overline{x})$.

Then $\overline{x}$ over-approximates $\lfp_{\basis}F$, that is,
$\lfp_{\basis}F\mathrel{\hyper\sqsubseteq} \hyper\gamma(\overline{x})$.
\myqed
\end{mythm}



Our goal is
over-approximation of 
the semantics of iteration of a loop-free $\Whiledt$ program $c$, 
relying on  Thm.~\ref{thm:hyperconcretization}.
Towards the goal, the next step
 is to find a suitable
$\overline{F}: \hyper{\overline{L}} \rightarrow \hyper{\overline{L}}$
that ``stepwise approximates'' $F=\sem{c}$, the collecting semantics of $c$.
  The next result implies that the $*$-transformation of 
$\semcp{\place}$ (defined in a usual manner in standard abstract interpretation, as mentioned in~\S{}\ref{subsec:abstinterp}) can be used in
 such $\overline{F}$.


 \begin{myprop}
 Let
$(L, \sqsubseteq), (\overline{L}, \overline{\sqsubseteq}), \gamma: \overline{L} \rightarrow L$ satisfy the hypotheses in Thm.~\ref{thm:hyperconcretization}.
 Assume that a continuous function $F: L \rightarrow L$ is stepwise approximated by a monotone function
  $\overline{F}: \overline{L}\rightarrow\overline{L}$, that is, 
$F\circ\gamma \sqsubseteq \gamma\circ\overline{F}$.
 Then  the *-continuous function $\hyper{F}: \hyper{L} \rightarrow \hyper{L}$ is over-approximated by the monotone and internal function $\hyper{\overline{F}}: \hyper{\overline{L}}\rightarrow \hyper{\overline{L}}$, i.e. $\hyper{F} \circ\hyper\gamma \mathrel{\hyper{\sqsubseteq}} \hyper{\gamma}\circ\hyper{\overline{F}}$.
 \myqed
 \end{myprop}


We summarize what we observed so far on nonstandard abstract
interpretation by instantiating the abstract domain to $\hyper{\CP_{n}}$.
In the following $\sem{c}$ is from Def.~\ref{def:whiledtsem}.
\begin{mycor}[soundness of nonstandard abstract interpretation on $\hyper{\CP_{n}}$]
\label{cor:soundnessOfHyperAbstractDomains}


 Let $c$ be a loop-free $\Whiledt$ command;
       and let $\basis\in\hyper{(\Pow(\R^{n}))}$
       and $\overline{x} \in \hyper{\CP_n}$ be such that $(\hyper{\semcp{c}})(\overline{x})
\mathrel{\hyper\sqsubseteq}
       \overline{x}$ and
       $\basis\mathrel{\hyper\sqsubseteq}\hyper{\gamma_{\CP_n}}(\overline{x})$.
        Then we have 
        $\lfp_{\basis}\sem{c}\mathrel{\hyper\sqsubseteq}\hyper{\gamma_{\CP_n}}(\overline{x})$.
\myqed
\end{mycor}

\subsection{Hyperwidening and Uniform Widening Operators}
\label{subsec:uniformWidening}
Towards our goal of using Thm.~\ref{thm:hyperconcretization}, the last
remaining step is to find a prefixed point $\overline{x}$,
i.e.\
 $\overline{F}(\overline{x})\mathrel{\hyper{\overline{\sqsubseteq}}}\overline{x}$.
This is where widening operators are standardly used; see~\S{}\ref{subsec:abstinterp}.


We can try $*$-transforming a (standard) notion---a strategy that we have used repeatedly in the current section. This
yields the following result, that has a problem that is discussed shortly.
\begin{mythm}
\label{thm:widenwithinf}
Let $(L, \sqsubseteq)$ be a preorder and $\nabla: L\times L\rightarrow L$ be a widening operator on $L$.
Let $F: \hyper{L} \rightarrow\hyper{L}$ 
be a monotone and internal function;
and $\basis\in \hyper{L}$ 
be such that $\basis \mathrel{\hyper\sqsubseteq} F(\basis)$.
 The  iteration \emph{hyper}-sequence 
$\langle X_{i}\rangle_{i\in\hyper\N}$---indexed by hypernaturals
 $i\in\hyper{\N}$---that is defined by
 $$
\footnotesize
\begin{array}{c}
X_0 = \basis,\quad   X_{i+1} =    \begin{cases}
    X_{i} & ({\text if }\; F(X_i) \mathrel{\hyper\sqsubseteq} X_{i})\\
    X_{i} \hyper\nabla F(X_{i}) & (otherwise)
   \end{cases}  {\rm for \; all}\; i \in \hyper\N
\end{array}
$$
reaches its limit within some hypernatural number of steps and the limit
 $\bigsqcup_{i\in\N}X_i$ is a prefixed point of $F$ such that $\basis
\mathrel{\hyper\sqsubseteq} \bigsqcup_{i\in\N}X_i$. \myqed
\end{mythm}
The problem of Thm.~\ref{thm:widenwithinf} is that the \emph{finite-step
 convergence} of iteration sequences for the original widening operator (described in Prop.~\ref{prop:widen})
 is now transferred to \emph{hyper\-finite-step convergence}. 
 This is not desired. All the entities from NSA that we have used so far
 are constructs in denotational semantics---whose only role is to ensure
 soundness of verification methodologies\footnote{Recall that $\Whiledt$
 is a \emph{modeling} language and we do not execute them.} and on which
 we never actually operate---and therefore their
 infinite/infinitesimal nature has been not a problem. In contrast,
 computation of the iteration hypersequence $\langle
 X_{i}\rangle_{i\in\hyper\N}$ is what we actually compute to
 over-approximate program semantics; and therefore its termination
 guarantee within $i\in\hyper{\N}$ steps (Thm.~\ref{thm:widenwithinf})
 is of no use.
 
 As a remedy we introduce a new notion of \emph{uniformity} of the (standard) widening
 operators. It strengthens the original termination condition
 (Def.~\ref{def:widen})
 by imposing a uniform bound $i$ for stability of arbitrary chains
 $\langle x_i \rangle \in L^\N$. Logically the change means
 replacing $\forall\exists$ by $\exists\forall$.

\begin{mydef}[uniform widening]\label{def:unifwiden}
 Let $(L, \sqsubseteq)$ be a preorder.
 A function $\nabla: L \times L \rightarrow L$ is said to be a \emph{uniform widening operator} if the following two conditions hold.
 \begin{itemize}
  \item (Covering) For any $x, y \in L$, $x \sqsubseteq x \nabla y$ and $y \sqsubseteq x \nabla y$.
  \item (Uniform termination) Let $x_0 \in L$. There exists a
	\emph{uniform bound} $i \in \N$ such that: for any ascending
	chain $\langle x_k \rangle \in L^\N$ starting from $x_0$, there
	exists $j \le i$ at which the chain  $\langle y_k \rangle \in
	L^\N$, defined by $y_0 = x_0$ and $y_{k+1} = y_{k} \nabla x_{k+1} {\rm\; for \; all}\; k \in \N$, stabilizes (i.e.\ $y_{j} = y_{j+1}$). 
 \end{itemize}
\end{mydef}
It is straightforward that uniform termination implies termination.

We investigate uniformity of some of the commonly-known widening operators on convex polyhedra.
\begin{mythm}\label{thm:uniformityOfKnownWidening}
 Among the three widening operators in~\S{}\ref{subsec:abstinterp}, $\nabla_{S}$ (Def.~\ref{def:stdwiden}) and $\nabla_{M}$ (Def.~\ref{def:widenupto}) are uniform, but $\nabla_{N}$ (\cite{Bagnara2005}) is not.
\myqed
\end{mythm}
For example, the widening operator $\nabla_{S}$ is uniform because once the first element $x_0$ of an iteration sequence is fixed, the length of the iteration sequence is at most the number of linear inequalities that define the convex polyhedra $x_0$.
However, $\nabla_{N}$ is not uniform because an iteration sequence can be arbitrarily long even if the first element of it is fixed,

The following theorem is a ``practical'' improvement of
Thm.~\ref{thm:widenwithinf};
its proof relies on instantiating the uniform bound $i$ 
 in a suitable $\LR$-formula with a Skolem
constant, before transfer.

\begin{mythm}
\label{thm:newunifwidenwithinfinitesimal}
 Let $(L, \sqsubseteq)$ be a preorder and $\nabla \in L \times L \rightarrow L$ be a uniform widening operator on $L$.
Let $F: \hyper{L} \rightarrow\hyper{L}$ 
be a monotone and internal function;
and $\basis\in L$ 
be such that $\hyper\basis \mathrel{\hyper\sqsubseteq} F(\hyper\basis)$.
 The  iteration sequence
 $\langle X_i\rangle_{i \in \N}$ defined by 
 $$
\footnotesize
\begin{array}{c}
X_0 = \hyper\basis,\quad 
    X_{i+1} = 
    \begin{cases}
     X_{i} & ({\text if }\; F(X_i) \mathrel{\hyper{\sqsubseteq}} X_{i})\\
     X_{i} \mathrel{\hyper{\nabla}} F(X_{i}) & (otherwise)
    \end{cases} \quad{\rm for \; all}\; i \in \N
\end{array}
$$
reaches its limit within some finite number of steps; and the limit
 $\bigsqcup_{i\in\N}X_i$ is a prefixed point of $F$ such that
 $\hyper\basis \mathrel{\hyper\sqsubseteq} \bigsqcup_{i\in\N}X_i$. \myqed
\end{mythm}

Note that uniformity of $\nabla$ is a \emph{sufficient condition} for
the termination of nonstandard iteration sequences (by
$\hyper{\nabla}$); Thm.~\ref{thm:newunifwidenwithinfinitesimal} does not
prohibit other useful widening operators in the nonstandard setting.
Furthermore, there can be a useful (nonstandard) widening operator
except for the ones $\hyper\nabla$ that arise
via standard ones $\nabla$.

 It is a direct consequence of Thm.~\ref{thm:newunifwidenwithinfinitesimal} and Thm.~\ref{thm:uniformityOfKnownWidening} that the analysis of $\Whiledt$ programs on $\hyper{\CP_n}$ is terminating with $\nabla_S$ or $\nabla_M$.

\section{Implementation and Experiments}\label{sec:implementation}
\subsection{Implementation}
We implemented a prototype tool for analysis of $\Whiledt$ programs.
The tool currently supports: $\hyper\CP_{n}$ as an abstract domain; and
$\hyper\nabla_{M}$, *-transformation of $\nabla_M$ in Def.~\ref{def:widenupto} as a widening operator.
Its input is 
a $\Whiledt$ program.
 It outputs a convex polyhedron that over-approximates the set of reachable memory states for each modes (or the values of discrete variables).
Our tool consists principally of the following two components:
1) an OCaml frontend for parsing, forming an iteration sequence and making the set $M$ for $\hyper\nabla_{M}$; and
2) a Mathematica backend for executing operations on convex polyhedra.
The two components are interconnected by a C++ program, via MathLink.

There are some libraries such as Parma Polyhedra Library~\cite{BagnaraHZ08SCP} that are commonly used to execute operations on convex polyhedra.
They cannot be used in our implementation because we have to handle the infinitesimal constant $\dt$ as an truly infinitesimal value.
Instead we implemented Chernikova's algorithm~\cite{Chernikova1964, Chernikova1965, Chernikova1968, LeVerge1992} symbolically, using \emph{computer algebra system (CAS)} on Mathematica based on Prop.~\ref{prop:dtToQE}.

Prop.~\ref{prop:dtToQE} ensures that the transformation from $\hyper A (\dt)$ \\to
$\exists r \in \R. \left(0<r \wedge \forall x \in \R. \left(0<x<r \Rightarrow A \left(x\right)\right)\right)$ does not violate the soundness of the analysis.
When we have to evaluate a formula including $\dt$, we instead resolve $\exists r \in \R. \left(0<r \wedge \forall x \in \R. \left(0<x<r \Rightarrow A \left(x\right)\right)\right)$ using CAS (e.g. quantifier elimination).


\subsection{Experiments}\label{subsec:experiments}
We analyzed two $\Whiledt$ programs---the water-level monitor (Fig.~\ref{fig:whileDtCodeCaseStudy}) and the thermostat
(Fig.~\ref{fig:thermostat})---with our prototype.
The experiments were on Apple MacBook Pro with 2.6 GHz Dual-core Intel Core i5 CPU and 8 GB memory and the execution times are the average of 10 runs.

\noindent
\textbf{Water-Level Monitor} 
This is a piecewise-linear dynamics
and a typical example used in hybrid automata literature. Our tool
automates the analysis presented in~\S{}\ref{sec:exampleOfAnalysis};
the execution time was 22.151 sec.

\noindent
\textbf{Thermostat} 
The dynamics of this example is beyond piecewise-linear.
The nonstandard abstract interpretation
successfully analyzes this example without explicit piece\-wise-linear
approximation. We believe this result witnesses a potential of our approach. We skip how it analyzes this example since the procedure is the same as the water-level monitor case.
Our tool executes in 2.259 sec.\ and outputs an approximation from which 
we obtain an invariant $18-54*\dt\leq x \leq 22+24*\dt$.

\section{Conclusions and Future Work}


We presented an extended abstract interpretation framework in which
hybrid systems are \emph{exactly} modeled as programs with infinitesimals.
 The logical infrastructure by \emph{nonstandard analysis} (in particular
the \emph{transfer principle}) establishes its 
soundness.
Termination is also ensured for \emph{uniform} widening operators.
Our prototype analyzer automates the extended abstract interpretation on the domain of convex polyhedra.

Regrettably our current implementation is premature and does not
compare---in precision or scalability---with the state-of-art tools for
hybrid system reachability such as  SpaceEx~\cite{Frehse11} and Flow*~\cite{ChenAS13}.
In fact the two examples in~\S{}\ref{subsec:experiments} are the only
ones that we have so far succeeded to analyze. For other
examples---especially nonlinear ones, to which our framework is
applicable in principle---the analysis results are too imprecise to be useful.
 To improve
there are some possible directions of future work to enhance the precision and scalability.
Firstly, we could utilize trace partitioning~\cite{Mauborgne2005}, narrowing operators (the use of narrowing operators in the domain of convex polyhedra is indicated in~\cite[\S{3.4}]{Henriksen2007}) and other techniques that have been introduced to enhance the precision of the analysis.
Secondly, 
we believe  abstract domains such as
\emph{ellipsoids}~\cite{Feret2004}, or some new ones that are tailored
to nonlinear dynamics, 
can improve our analyzer.
Finally, 
the lack of scalability is mainly due to our current way of eliminating $\dt$ (namely via
Prop.~\ref{prop:dtToQE}): it relies on \emph{quantifier elimination
(QE)} that is highly expensive. A faster alternative is desired.



\bibliographystyle{splncs03} 
\bibliography{../../../library}

\begin{thebibliography}{10}
\providecommand{\url}[1]{\texttt{#1}}
\providecommand{\urlprefix}{URL }

\bibitem{Alur1992}
Alur, R., Courcoubetis, C., Henzinger, T.A., Ho, P.: Hybrid automata: An
  algorithmic approach to the specification and verification of hybrid systems.
  In: Hybrid Systems. pp. 209--229 (1992)

\bibitem{BagnaraHZ08SCP}
Bagnara, R., Hill, P.M., Zaffanella, E.: The {Parma Polyhedra Library}: Toward
  a complete set of numerical abstractions for the analysis and verification of
  hardware and software systems. Science of Computer Programming  72(1--2),
  3--21 (2008)

\bibitem{Bagnara2005}
Bagnara, R., Hill, P.M., Ricci, E., Zaffanella, E.: Precise widening operators
  for convex polyhedra. Sci. Comput. Program.  58(1-2),  28--56 (2005)

\bibitem{BeauxisM11}
Beauxis, R., Mimram, S.: A non-standard semantics for {Kahn} networks in
  continuous time. In: CSL. pp. 35--50 (2011)

\bibitem{ChenAS13}
Chen, X., {\'{A}}brah{\'{a}}m, E., Sankaranarayanan, S.: Flow*: An analyzer for
  non-linear hybrid systems. In: Computer Aided Verification - 25th
  International Conference, {CAV} 2013, Saint Petersburg, Russia, July 13-19,
  2013. Proceedings. pp. 258--263 (2013)

\bibitem{Chernikova1964}
Chernikova, N.: Algorithm for finding a general formula for the non-negative
  solutions of a system of linear equations. USSR Computational Mathematics and
  Mathematical Physics  4(4),  151--158 (1964)

\bibitem{Chernikova1965}
Chernikova, N.: Algorithm for finding a general formula for the non-negative
  solutions of a system of linear inequalities. USSR Computational Mathematics
  and Mathematical Physics  5(2),  228--233 (1965)

\bibitem{Chernikova1968}
Chernikova, N.: Algorithm for discovering the set of all the solutions of a
  linear programming problem. USSR Computational Mathematics and Mathematical
  Physics  8(6),  282--293 (1968)

\bibitem{Cousot1981}
Cousot, P.: Semantic foundations of program analysis. In: Muchnick, S., Jones,
  N. (eds.) Program Flow Analysis: Theory and Applications, chap.~10, pp.
  303--342. Prentice-Hall, Inc{.}, Englewood Cliffs, New Jersey (1981)

\bibitem{Cousot1977}
Cousot, P., Cousot, R.: Abstract interpretation: {A} unified lattice model for
  static analysis of programs by construction or approximation of fixpoints.
  In: Conference Record of the Fourth {ACM} Symposium on Principles of
  Programming Languages, Los Angeles, California, USA, January 1977. pp.
  238--252 (1977)

\bibitem{Cousot1992a}
Cousot, P., Cousot, R.: Abstract interpretation frameworks. J. Log. Comput.
  2(4),  511--547 (1992)

\bibitem{Cousot2005}
Cousot, P., Cousot, R., Feret, J., Mauborgne, L., Min{\'{e}}, A., Monniaux, D.,
  Rival, X.: The astre{\'{e}} analyzer. In: Programming Languages and Systems,
  14th European Symposium on Programming,ESOP 2005, Held as Part of the Joint
  European Conferences on Theory and Practice of Software, {ETAPS} 2005,
  Edinburgh, UK, April 4-8, 2005, Proceedings. pp. 21--30 (2005)

\bibitem{Cousot1978}
Cousot, P., Halbwachs, N.: Automatic discovery of linear restraints among
  variables of a program. In: Conference Record of the Fifth Annual {ACM}
  Symposium on Principles of Programming Languages, Tucson, Arizona, USA,
  January 1978. pp. 84--96 (1978)

\bibitem{Feret2004}
Feret, J.: Static analysis of digital filters. In: Programming Languages and
  Systems, 13th European Symposium on Programming, {ESOP} 2004, Held as Part of
  the Joint European Conferences on Theory and Practice of Software, {ETAPS}
  2004, Barcelona, Spain, March 29 - April 2, 2004, Proceedings. pp. 33--48
  (2004)

\bibitem{Franzle2007}
Fr{\"{a}}nzle, M., Herde, C., Teige, T., Ratschan, S., Schubert, T.: Efficient
  solving of large non-linear arithmetic constraint systems with complex
  boolean structure. {JSAT}  1(3-4),  209--236 (2007)

\bibitem{Frehse05}
Frehse, G.: Phaver: Algorithmic verification of hybrid systems past hytech. In:
  Hybrid Systems: Computation and Control, 8th International Workshop, {HSCC}
  2005, Zurich, Switzerland, March 9-11, 2005, Proceedings. pp. 258--273 (2005)

\bibitem{Frehse11}
Frehse, G., Guernic, C.L., Donz{\'{e}}, A., Cotton, S., Ray, R., Lebeltel, O.,
  Ripado, R., Girard, A., Dang, T., Maler, O.: Spaceex: Scalable verification
  of hybrid systems. In: Computer Aided Verification - 23rd International
  Conference, {CAV} 2011, Snowbird, UT, USA, July 14-20, 2011. Proceedings. pp.
  379--395 (2011)

\bibitem{Goldblatt1998}
Goldblatt, R.: Lectures on the Hyperreals: An Introduction to Nonstandard
  Analysis. Graduate Texts in Mathematics, Springer New York (1998)

\bibitem{Halbwachs1979}
Halbwachs, N.: D^^c3^^a9termination automatique de relations lin^^c3^^a9aires
  v^^c3^^a9rifi^^c3^^a9es par les variables d'un programme. Th^^c3^^a8se de 3e
  cycle, Universit^^c3^^a9 Scientifique et M^^c3^^a9dicale de Grenoble (1979)

\bibitem{Halbwachs1993}
Halbwachs, N.: Delay analysis in synchronous programs. In: Computer Aided
  Verification, 5th International Conference, {CAV} '93, Elounda, Greece, June
  28 - July 1, 1993, Proceedings. pp. 333--346 (1993)

\bibitem{Halbwachs1997}
Halbwachs, N., Proy, Y., Roumanoff, P.: Verification of real-time systems using
  linear relation analysis. Formal Methods in System Design  11(2),  157--185
  (1997)

\bibitem{Hasuo2012}
Hasuo, I., Suenaga, K.: Exercises in nonstandard static analysis of hybrid
  systems. In: Computer Aided Verification - 24th International Conference,
  {CAV} 2012, Berkeley, CA, USA, July 7-13, 2012 Proceedings. pp. 462--478
  (2012)

\bibitem{Henriksen2007}
Henriksen, K.S., Banda, G., Gallagher, J.P.: Experiments with a convex
  polyhedral analysis tool for logic programs. CoRR  abs/0712.2737 (2007),
  \url{http://arxiv.org/abs/0712.2737}

\bibitem{Henzinger95}
Henzinger, T.A., Ho, P.: Algorithmic analysis of nonlinear hybrid systems. In:
  Computer Aided Verification, 7th International Conference, Li{\`{e}}ge,
  Belgium, July, 3-5, 1995, Proceedings. pp. 225--238 (1995)

\bibitem{Henzinger1997}
Henzinger, T.A., Ho, P., Wong{-}Toi, H.: {HYTECH:} {A} model checker for hybrid
  systems. {STTT}  1(1-2),  110--122 (1997)

\bibitem{Hurd1985}
Hurd, A., Loeb, P.: An Introduction to Nonstandard Real Analysis. Pure and
  Applied Mathematics, Elsevier Science (1985)

\bibitem{Kido2013}
Kido, K.: An Alternative Denotational Semantics for an Imperative Language with
  Infinitesimals. Bachelor's thesis, The University of Tokyo: Japan (2013)

\bibitem{NSAI}
Kido, K., Chaudhuri, S., Hasuo, I.: Source code of the prototype nonstandard
  abstract interpreter (2015), \url{http://www-mmm.is.s.u-tokyo.ac.jp/~kkido/}

\bibitem{LeVerge1992}
Le~Verge, H.: {A note on Chernikova's Algorithm}. Tech. Rep. 635, IRISA,
  Rennes, France (Feb 1992)

\bibitem{Mauborgne2005}
Mauborgne, L., Rival, X.: Trace partitioning in abstract interpretation based
  static analyzers. In: Programming Languages and Systems, 14th European
  Symposium on Programming,ESOP 2005, Held as Part of the Joint European
  Conferences on Theory and Practice of Software, {ETAPS} 2005, Edinburgh, UK,
  April 4-8, 2005, Proceedings. pp. 5--20 (2005)

\bibitem{PlatzerQ08}
Platzer, A., Quesel, J.D.: {KeYmaera}: A hybrid theorem prover for hybrid
  systems. In: Armando, A., Baumgartner, P., Dowek, G. (eds.) IJCAR. LNCS, vol.
  5195, pp. 171--178. Springer (2008)

\bibitem{Robinson1966}
Robinson, A.: Non-standard Analysis. Studies in logic and the foundations of
  mathematics, North-Holland Pub. Co. (1966)

\bibitem{Suenaga2011}
Suenaga, K., Hasuo, I.: Programming with infinitesimals: {A} while-language for
  hybrid system modeling. In: Automata, Languages and Programming - 38th
  International Colloquium, {ICALP} 2011, Zurich, Switzerland, July 4-8, 2011,
  Proceedings, Part {II}. pp. 392--403 (2011)

\bibitem{Suenaga2013}
Suenaga, K., Sekine, H., Hasuo, I.: Hyperstream processing systems: nonstandard
  modeling of continuous-time signals. In: The 40th Annual {ACM}
  {SIGPLAN-SIGACT} Symposium on Principles of Programming Languages, {POPL}
  '13, Rome, Italy - January 23 - 25, 2013. pp. 417--430 (2013)

\bibitem{Winskel1993}
Winskel, G.: The Formal Semantics of Programming Languages: An Introduction.
  MIT Press, Cambridge, MA, USA (1993)

\end{thebibliography}
  
\newpage
\appendix
\section{Further on NSA in Superstructure}
\label{appendix:NSAPrimer}
The definitions and results  listed below are all well-established and commonly
used in NSA. We follow~\cite[Chap.~II]{Hurd1985}, in which
more details can be found.

\begin{myrem}[choice of the index set $I$]\label{remark:choiceOfNatAsI}
 In~\S\ref{subsec:preliminariesNSA} we used the set $\N$ of natural
 numbers as the index set $I$. 
 It is common in NSA, however, to use $I$ that is bigger than
 $\N$, and an ultrafilter $\filt\subseteq\pow (I)$ over $I$. The merit of
 doing so is that the resulting monomorphism $\hyper{(\place)}$ (see below) can be
 chosen to be an \emph{enlargement}; see~\cite[Chap.~II]{Hurd1985}.
In what follows, however, we favor concreteness and keep using $I=\N$
as the index set. 
\end{myrem}


The transfer principle is a powerful result and we  rely on
it in the subsequent developments. Here are the first
examples of its use; they are proved by transferring a suitable formula $A$.

\begin{mylem}\label{lemma:firstUseOfTransfer}
\begin{enumerate}
 \item\label{item:*transformRestricted}
   For $a\in V(\baseSet)\setminus \baseSet$ we obtain an injective map
  \begin{equation}\label{equation:*TransferMapRestricted}
 \footnotesize
\begin{array}{l}
    \hyper{(\place)}
  \;:\; a\longrightarrow\hyper{a}\enspace,\quad
  (b\in a)\longmapsto (\hyper{b}\in\hyper{a})
\end{array}  
\end{equation}
 as a restriction of $\hyper{(\place)}$ in~(\ref{equation:*TransferMap}).
 \item\label{item:starFiniteIsFinite} If  $a$ is a finite set, the 
      map~(\ref{equation:*TransferMapRestricted}) is an isomorphism
      $a\iso\hyper{a}$.
 \item \label{item:functionSpaceTransformed} Let $a\to b$ be the
       set of functions from $a$ to $b$. We have
       \begin{math}
	\hyper{(a\to b)}
	\subseteq
	\hyper{a}\to\hyper{b}
       \end{math}.
 \item\label{item:productUnionTransformed}
      $\hyper{(a_{1}\times\cdots\times a_{m})}
      = \hyper{a_{1}} \times\cdots\times \hyper{a_{m}}
    $; and  
      $\hyper{(a_{1}\cup\cdots\cup a_{m})}
      = \hyper{a_{1}} \cup\cdots\cup \hyper{a_{m}}
    $.
 \item\label{item:binaryRelTransformed} For a binary relation $r\subseteq a\times a$, we have
       $\hyper{r}\subseteq \hyper{a}\times\hyper{a}$. Moreover,  $r$ is
      an order if and only if $\hyper{r}$ is an order. \myqed
\end{enumerate}
\end{mylem}

\paragraph{Internal Sets}
The distinction between \emph{internal} and \emph{external} entities is
central in NSA.  In this paper however it is much of formality, since
all the entities we use are internal.  Here we present only the relevant
definitions, leaving their intuitions to~\cite[\S{}II.6]{Hurd1985}.
 In Appendix~\ref{appendix:domainTheoryTransferred}, 
 especially
 Rem.~\ref{remark:significanceOfInternal}, we will see that being
 internal is
  crucial 
 for transfer.

\begin{mydef}[internal entity]\label{definition:internalEntity}
 An element $b\in V(\hyper{\baseSet})$ is \emph{internal} with respect to
 $\hyper{(\place)}:V(\baseSet)\to V(\hyper{\baseSet})$ if there is $a\in V(\baseSet)$ such
 that $b\in \hyper{a}$. It is \emph{external} if it is not internal.
\end{mydef}

\begin{mylem}\label{lemma:charInternalFunction}
A function $f:\hyper{a}\to\hyper{b}$ is internal if and only if
 $f\in\hyper{(a\to b)}$. \myqed
\end{mylem}

\paragraph{The Ultrapower Construction}
We collect some necessary facts about
the ultrapower construction of the monomorphism $\hyper{(\place)}$
in~(\ref{equation:*TransferMap}). Its details are beyond our scope; they 
are found in~\cite[\S{}II.4]{Hurd1985}.

The map  $\hyper{(\place)}$ in fact factorizes into the following three
steps.
\begin{equation}\label{diagram:monomorphismFactorized}\footnotesize
 \vcenter{\xymatrix@R=.6em{
 {V(\baseSet)}
   \ar[r]^-{\hyper{(\place)}}
   \ar[d]_-{\overline{(\place)}}
 &
 {V(\hyper{\baseSet})}
 \\
 {\bigcup_{n\in\N}\bigl(V_{n}(\baseSet)\setminus V_{n-1}(\baseSet)\bigr)^{I}}
    \ar[r]_-{[\place]}
 &
 {\prod^{0}_{\filt} V(\baseSet)}
    \ar[u]_{M}
}}
\end{equation}
The first factor $\overline{(\place)}$ maps $a\in V(\baseSet)$ to the constant
function $\overline{a}$ such that $\overline{a}(i)=a$ for each $i\in I$;
recall that we have chosen $I=\N$ (Rem.~\ref{remark:choiceOfNatAsI}). The second
 $[\place]$ takes a quotient modulo the ultrafilter $\filt$;
finally the third factor $M$ is the so-called \emph{Mostowski collapse}.

For an intuition let us exhibit these maps in the simple setting of~\S{}\ref{subsec:preliminariesNSA}.
The first factor  $\overline{(\place)}$ corresponds to forming constant streams:
$a\mapsto \overline{a}=(a,a,\dotsc)$. The second $[\place]$ is quotienting
modulo $\sim_{\filt}$ of~(\ref{equation:defOfSimFilt}). The third map
$M$ does nothing---it is a book-keeping function
that is only needed in the extended setting of superstructures.

The next result \cite[Thm.~4.5]{Hurd1985} is about ``starting from the lower-left corner'' in~(\ref{diagram:monomorphismFactorized}). It
follows from the definition of $M$ and  is a crucial
step in the proof of the transfer principle (Lem.~\ref{lemma:transferPrinciple}). It serves as an important
lemma, too,  later for
the semantics of $\Whiledt$.
\begin{mylem}[\L{}o\'{s}' theorem]\label{lemma:sequenceAsHyperEntity}
Let $A$ be a formula in $\LX$ with its free variables contained in $\{x_{1},\dotsc,
 x_{m}\}$; and $a_{1},\dotsc,a_{m}\in
 {\bigcup_{n\in\N}\bigl(V_{n}(\baseSet)\setminus V_{n-1}(\baseSet)\bigr)^{I}}$. Then
 \begin{displaymath}
\begin{array}{l}
 \hyper{A}\bigl[\,M[a_{1}]/x_{1},\dotsc, M[a_{m}]/x_{m}\,\bigr]
  \;\text{is valid}
 \\
 \Longleftrightarrow\quad
 \bigl\{ i\in I\,\mid\;
  A[a_{1}(i)/x_{1},\dotsc, a_{m}(i)/x_{m}]
  \;\text{is valid}
 \bigr\} \in \filt\enspace.
\end{array} 
\end{displaymath}
 As a special case, let $S\in V(\baseSet)$, then
 \begin{equation}
\begin{array}{l}
   M[a] \in \hyper{S}
  \quad\Longleftrightarrow\quad
  a(i)\in S\;\;\text{for almost every $i$.}
\end{array}  
\tag*{\myqed}
 \end{equation}
\end{mylem}

\begin{mycor}\label{corollary:MFuncMArg}
  Let $a,b\in V(\baseSet)$; and for each $i\in I$, $f_{i}\in (a\to b)$ and
 $x_{i}\in a$. Then $M[(f_{i})_{i\in I}]$ is an internal function
	$\hyper{a}\to\hyper{b}$; and $M[(x_{i})_{i\in I}] \in
	\hyper{a}$. Moreover, 
 \begin{equation}
\begin{array}{l}
   M[(f_{i}(x_{i}))_{i\in I}]
 =
 \Bigl(  M[(f_{i})_{i\in I}]
 \Bigr)
 \Bigl(
 M[(x_{i})_{i\in I}] 
 \Bigr)\enspace.
\end{array} 
\tag*{\myqed}
  \end{equation}
\end{mycor}

\section{Appendix: Domain Theory, Transferred}\label{appendix:domainTheoryTransferred}
The collecting semantics of $\Whiledt$ is introduced by solving recursive equations on $\hyper\Pow(\R^n)$.
Here we present
necessary theoretical foundations---they are  like
in~\cite[\S{}2.2]{BeauxisM11} and~\cite{Suenaga2013}---identifying the
set $\hyper{\Pow(\R^n)}$ as a hyperdomain and *-transferring domain theory.

 The current section is an adaptation is what appeared in the appendix
 of~\cite{Suenaga2013}; and 
 the definitions and results  are similar to those
 in~\cite[\S{}2.2]{BeauxisM11}, where what we call a hyperdomain is
 called an \emph{internal domain}, and a *-continuous function is called
 an
 \emph{internal continuous function}. The way we formulate these notions
 is however a bit different: we favor more explicit use of *-transforms, since
 this aids deductive verification via the transfer principle.


\begin{mydef}\label{definition:baseSetOfCurrentPaper}
 In what follows we employ the theory of NSA presented
 in Appendix~\ref{appendix:NSAPrimer}. As
 the base set of a superstructure $V(X)$
 (Def.~\ref{definition:superstructure}), we take
 $X=\R\cup\B\cup\Var$.
\end{mydef}

\begin{mydef}[hyperdomain]\label{definition:hyperdomain}
 A \emph{hyperdomain} is  the pair of *-transforms
 $(\hyper{D},\hyper{\sqsubseteq})$ of a cpo $(D,\sqsubseteq)$. 
\end{mydef}
\begin{myexpl}\label{ex:hyperdomainInCurrentWork}
 The set $\pow(\Var\to\R)$ is a complete lattice with respect to the
 inclusion order $\subseteq$, therefore is a cpo. Its $*$-transfer 
 $
\bigl(\,
\hyper{\bigl(\pow(\Var\to\R)\bigr)},\,
\hyper{\subseteq}
\,\bigr)
$ constitutes a hyperdomain. 

We note that the set $\hyper{\bigl(\pow(\Var\to\R)\bigr)}$ coincides
 with the set of internal subsets of the space $\{f\colon
 \hyper{\Var}\to\hyper{\R}\mid f\text{ is an internal function}\}$.
 Moreover, under the assumption that $\Var$ is a finite set (e.g.\ the
 set of variables occurring in a program $c$), we can see that the last
 set
$\{f\colon
 \hyper{\Var}\to\hyper{\R}\mid f\text{ is an internal function}\}$
 coincides with the function space $\Var\to\hyper{\R}$. For this we use 
Lem.~\ref{lemma:firstUseOfTransfer}.\ref{item:productUnionTransformed}.
\end{myexpl}

 Note that $\hyper{\sqsubseteq}$ is an order in $\hyper{D}$
 (Lem.~\ref{lemma:firstUseOfTransfer}.\ref{item:binaryRelTransformed}).
 Hyperdomain is the notion on which we wish to establish a suitable fixed point
 property.\footnote{We believe an even more general setting is possible, by defining a hyperdomain to be an internal set $D'\in V(\hyper{X})$
 that satisfies a suitable formula like
 $\mathsf{CPO}_{a,r}$ in~(\ref{equation:CPOAndContinuityInLX}). Here we
 do not need such generality.}  Towards that goal, we first formulate
 the definitions of cpo and continuous function as
 $\LX$-formulas, so that they can be transferred.
\begin{displaymath}
 \begin{array}{l}
  \mathsf{BinRel}_{a,r}
  :\equiv\;
  r\subseteq a\times a
\qquad
 \mathsf{Refl}_{a,r}
  :\equiv\;
 \forall x\in a.\, \rtuple{x,x}\in r
\\
 \mathsf{Trans}_{a,r}
  :\equiv\;
 \forall x,y,z\in a.\, 
\bigl(\,
\rtuple{x,y}\in r
 \land
\rtuple{y,z}\in r
 \Rightarrow
\rtuple{x,z}\in r
\,\bigr)
\\
 \mathsf{AntiSym}_{a,r}
  :\equiv\;
 \forall x,y\in a.\, 
\bigl(\,
\rtuple{x,y}\in r
 \land
\rtuple{y,x}\in r
 \Rightarrow
 x =y
\,\bigr)
\\
 \mathsf{Poset}_{a,r}
  :\equiv
  \;
  \mathsf{BinRel}_{a,r} \land
  \mathsf{Refl}_{a,r} \land
  \mathsf{Trans}_{a,r} \land
  \mathsf{AntiSym}_{a,r} 
\\
 \mathsf{HasBot}_{a,r}
 :\equiv\;
 \exists x\in a.\, \forall y\in a.\, \rtuple{x,y}\in r
\\
 \mathsf{AscCn}_{a,r}(s)
 :\equiv\;
 \forall x,x'\in \N.\, (x\le x'\Rightarrow \rtuple{s(x),s(x')}\in r)
\\
 \mathsf{UpBd}_{a,r}(b,s)
 :\equiv\;
 \forall x\in \N.\, (\rtuple{s(x),b}\in r)
\\
 \mathsf{Sup}_{a,r}(p,s)
 :\equiv\;
 \mathsf{UpBd}_{a,r}(p,s) \land
 \forall b\in a.\, ( \mathsf{UpBd}_{a,r}(b,s) \Rightarrow \rtuple{p,b}\in
 r)
  \end{array}
 \end{displaymath}
 Recall that the inclusion $\N\subseteq X$ is assumed
 (Def.~\ref{definition:superstructure}).  These formulas are used in:
 \begin{equation}\label{equation:CPOAndContinuityInLX}
  \begin{array}{rl}
  \mathsf{CPO}_{a,r}
 :\equiv\;&
  \mathsf{Poset}_{a,r}\land\mathsf{HasBot}_{a,r}\land
\\
\qquad&
  \forall s\in (\N\to a).\, 
  \bigl(\,
  \mathsf{AscCn}_{a,r}(s) \Rightarrow\exists p\in a.\, \mathsf{Sup}_{a,r}(p,s)
\,\bigr)\enspace,
  \\
  \mathsf{Conti}_{a_{1},r_{1},a_{2},r_{2}}(f)
 :\equiv\;&
  \forall s\in (\N\to a_{1}).\, 
  \forall p\in  a_{1}.\, 
  \\
\qquad&
\bigl(\,
  \mathsf{AscCn}_{a_{1},r_{1}}(s) \land \mathsf{Sup}_{a_{1},r_{1}}(p,s)
  \Rightarrow \mathsf{Sup}_{a_{2},r_{2}}(f(p),f\co s)
\,\bigr)\enspace. \end{array}
\end{equation}

\auxproof{
*** This notion is problematic since the formula
$\mathsf{CPO}$ cannot *-transformed easily.  In particular,
 the quantifier $s\in(\N\to a)$ is a syntax sugar and we do not know 
 how to *-transform it ****
\begin{mydef}[*-Cpo]\label{definition:*cpo}
Let $a\in
 V(\hyper{X})$ and $r$ be an order on $a$. The pair $(a,r)$ is
 said to be
 a \emph{*-cpo} in $V(\hyper{X})$ if the formula
 $\hyper{\mathsf{CPO}}(a,r)$
 is valid in $V(\hyper{X})$.
\end{mydef}

The *-transform of the prefix order $\sqsubseteq$ on $\C^{\infty}$
(Lem.~\ref{lemma:streamsFormCpo}) 
is an order $\hyper{\sqsubseteq}$
between hyperstreams (Lem.~\ref{lemma:firstUseOfTransfer}.\ref{item:binaryRelTransformed}).
\begin{mylem}\label{lemma:hyperstreamForm*CPO}
 The set $\CHypSt$ of hyperstreams, together with
 $\hyper{\sqsubseteq}$, constitutes a *-cpo in $V(\hyper{X})$.
\end{mylem}
\begin{myproof}
 The formula $\mathsf{CPO}(\C^{\infty}, \sqsubseteq)$ is closed and
 valid in $V(X)$ (Lem.~\ref{lemma:streamsFormCpo}). Thus by transfer,
 we have $\models \hyper{\mathsf{CPO}}(\CHypSt, \hyper{\sqsubseteq})$.
\myqed
\end{myproof}
}

\begin{mydef}[*-continuous function]\label{definition:*contiFunc}
 Let $(\hyper{D_{1}},\hyper{\sqsubseteq_{1}})$ and
$(\hyper{D_{2}},\hyper{\sqsubseteq_{2}})$ 
be  hyperdomains. A function
 $f:\hyper{D_{1}}\to\hyper{D_{2}}$ is
 \emph{*-continuous} if it is internal and satisfies the *-transform of the formula
 $\mathsf{Conti}_{D_{1},\sqsubseteq_{1},D_{2},\sqsubseteq_{2}}$. That
 is 
 to be precise:
 \begin{math}
   \hyper{(\mathsf{Conti}_{D_{1},\sqsubseteq_{1},D_{2},\sqsubseteq_{2}})}(f)
 \end{math} is valid.\footnote{We note that the condition is different from
 (somewhat informal) ``$\hyper{\mathsf{Conti}}_{
  \hyper{D_{1}},\hyper{\sqsubseteq_{1}},
  \hyper{D_{2}},\hyper{\sqsubseteq_{2}}
}(f)$ is valid.'' In the former a chain $s$ ranges over internal functions $s\in
 \hyper{(\N\to D_{1})}$, while in the latter $s$ can also be an external
 function $\hyper{\N}\to\hyper{D_{1}}$.}
 The set of *-continuous functions from $\hyper{D_{1}}$
to $\hyper{D_{2}}$ is denoted by $\hyper{D_{1}}\scto\hyper{D_{2}}$.
\end{mydef}

\begin{mylem}\label{lemma:*contiFuncSpAsStarTransform}
 $(\hyper{D_{1}}\scto\hyper{D_{2}})=\hyper{(D_{1}\cto D_{2})}$. Here
 $\cto$ denotes the set of continuous functions. 
\end{mylem}
\begin{myproof}
 Assume $f\in\hyper{(D_{1}\cto D_{2})}$. The following closed formula is
 valid in $V(X)$:
 \begin{displaymath}
  \forall f'\in (D_{1}\to D_{2}).\, 
\bigl( f'\in(D_{1}\cto D_{2})
 \Leftrightarrow
\mathsf{Conti}(f')
\bigr)\enspace,
 \end{displaymath}
 where $\mathsf{Conti}$ is short for
 $\mathsf{Conti}_{D_{1},\sqsubseteq_{1},D_{2},\sqsubseteq_{2}}$. By
 transfer we have
 \begin{equation}\label{equation:1:lemma:*contiFuncSpAsStarTransform}
  \forall f'\in \hyper{(D_{1}\to D_{2})}.\, 
\bigl( f'\in\hyper{(D_{1}\cto D_{2})}
 \Leftrightarrow
\hyper{\mathsf{Conti}}(f')
\bigr)
 \end{equation}
 valid in $V(\hyper{X})$. Thus $f$ satisfies
 $\hyper{\mathsf{Conti}}(f')$.  Obviously $f$ is internal;
 therefore
 $f\in(\hyper{D_{1}}\scto\hyper{D_{2}})$.

 Conversely, assume $f\in(\hyper{D_{1}}\scto\hyper{D_{2}})$. By the
 definition of *-continuity, $f$ is internal, hence by
 Lem.~\ref{lemma:charInternalFunction} we have $f\in\hyper{(D_{1}\to
 D_{2})}$.
 Moreover, using the definition of *-continuity and~(\ref{equation:1:lemma:*contiFuncSpAsStarTransform}), we have
 $f\in\hyper{(D_{1}\cto D_{2})}$. \myqed
\end{myproof}

\begin{mylem}\label{lemma:fixedPtInHyperdomain}
 Let $(\hyper{D},\hyper{\sqsubseteq})$ be a hyperdomain. Then a
 *-continuous function $f:\hyper{D}\to\hyper{D}$
 has a least
 fixed point. Moreover, the function $\hyper{\mu}$ that maps $f$ to its least fixed
 point $(\hyper{\mu})(f)$ is *-continuous.
\auxproof{it is moreover described as the supremum
 \begin{math}
  \dsup_{n\in\hyper{\N}}f^{n}(\bot)
 \end{math}.
}
\end{mylem}
\begin{myproof}
By the usual construction in a cpo, we obtain the map
\begin{displaymath}
 \begin{array}{l}
  \mu\;:\;(D\cto D)\cto D\enspace,\quad
  f\mapsto \dsup_{n\in\N}f^{n}(\bot)\enspace.
 \end{array}
\end{displaymath}
Continuity of $\mu$ is easy and standard.
As its *-transform
 we obtain a function
\begin{math}
  \hyper{\mu}:(\hyper{D}\scto \hyper{D})\scto \hyper{D}
\end{math}, where we used Lem.~\ref{lemma:*contiFuncSpAsStarTransform}
 and~\ref{lemma:firstUseOfTransfer}. 
 The fact that $\hyper{\mu}$ returns least fixed points is shown by the
 transfer
of the following $\LX$-formula.
 \begin{multline*}
    \forall f\in (D\cto D). \, 
 \bigl(\;
 f(\mu(f)) = \mu (f) \;\land
 \forall x\in D.\, (f(x)=x\;\Rightarrow \mu(f  )\sqsubseteq x)
 \;\bigr)
\tag*{\myqed}
\end{multline*}
\auxproof{*** obsolete ***
 The  formula $\mathsf{FixPt}$ below expresses the construction of  least
 fixed points in the cpo $D$. It is closed and valid in $V(X)$.
 \begin{displaymath}\label{equation:fixedPtInLX}\footnotesize
  \begin{array}{l}
  \mathsf{FixPt}  :\equiv
\\
 \forall f\in (D\to D).\,
   \Bigl(
     \mathsf{Conti}_{D,\sqsubseteq,D,\sqsubseteq}(f)
    \Rightarrow
 \exists s:(\N\to D).\, 
\exists p\in D.\,
 \\\quad
    s(0)=\bot \land 
\forall n\in \N.\, 
   \bigl(
  s(n+1)=f(s(n))    
\bigr)
  \land
\\\quad
 \mathsf{Sup}_{D,\sqsubseteq}(p,s)
  \land f(p)=p
  \land \forall y\in D.\, 
  \bigl(f(y)=y\;\Rightarrow p\sqsubseteq y\bigr)
\Bigr)
  \end{array}
 \end{displaymath}
 Thus its *-transform $\hyper{\mathsf{FixPt}}$ is valid by transfer.  

By *-continuity, 
the given function 
 $f:\hyper{D}\to\hyper{D}$ 
is internal,  hence $f\in\hyper{(D\to D)}$
 (Lem.~\ref{lemma:charInternalFunction}); 
moreover it satisfies
 $ \hyper{(\mathsf{Conti}_{D,\sqsubseteq,D,\sqsubseteq})}(f)$. The rest
 of the formula $\hyper{\mathsf{FixPt}}$ states that $p$ is the least
 fixed point of $f$, and that it is the supremum of the internal
 hyperfunction $s:\hyper{\N}\to\hyper{D}$, defined by
 \begin{displaymath}
  s(0):=\bot_{\hyper{D}}\enspace\quad\text{and}\quad
  s(n+1):= s(n) \quad\text{for $\forall n\in \hyper{\N}$}
 \end{displaymath}
 (whose well-definedness is easily seen by transfer). \myqed
}
\end{myproof}

\begin{myrem}\label{remark:significanceOfInternal}
It is crucial  in the previous lemma that
 $f:\hyper{D}\to\hyper{D}$ is an internal function. Specifically: recall that
 a formula $A$ must be closed in order to be transferred
 (Lem.~\ref{lemma:transferPrinciple}); and that
in $\LX$ only bounded quantifiers ($\forall x\in s$ with some bound $s$)
are allowed. For internal $f$ we find such a bound by $f\in\hyper{(D\to
 D)}$; for external $f$ this is not possible.

\end{myrem}

\section{Appendix: Omitted Proofs}\label{appendix:omittedProofs}

\subsection{Proof of Thm.~\ref{prop:dtToQE}}
\begin{myproof}
Assume that $$0<r \wedge \forall x \in \R. \left(0<x<r \Rightarrow A \left(x\right)\right)$$ is valid for some $r \in \R$.
By transfer, $$0<r \wedge \forall x \in \hyper\R. \left(0<x<r \Rightarrow\hyper A \left(x\right)\right)$$ is also valid for that $r$.
This implies $\hyper A (\dt)$ since $0 < \dt < r$ for any positive $r\in\R$.
\myqed
\end{myproof}

Hereafter in the proofs we use the following $\LU$-formulas.

\begin{mydef}\label{def:lunotations}
We define the following $\LU$-formulas:
 \begin{align*}
  &\mathsf{Refl}_{L, \sqsubseteq} :\equiv\; \forall l \in L.\; (l, l) \in \sqsubseteq \\
  &\mathsf{Trans}_{L, \sqsubseteq} :\equiv\; \forall l, m, n \in L.\; \Bigl(\bigl((l, m) \in {\sqsubseteq} \wedge (m, n) \in \sqsubseteq\bigr) \Rightarrow (l, m) \in {\sqsubseteq}\Bigr) \\
  &\mathsf{AntiSym}_{L, \sqsubseteq} :\equiv\; \forall l, m \in L.\; \Bigl(\bigl((l, m) \in \sqsubseteq \wedge (m, l) \in {\sqsubseteq} \bigr) \Rightarrow l = m \Bigr)\\
  &\Preord_{L, \sqsubseteq} :\equiv\; \mathsf{Refl}_{L, \sqsubseteq} \wedge \mathsf{Trans}_{L, \sqsubseteq} \\
  &\Poset_{L, \sqsubseteq} :\equiv\; \mathsf{Refl}_{L, \sqsubseteq} \wedge \mathsf{Trans}_{L, \sqsubseteq} \wedge \mathsf{AntiSym}_{L, \sqsubseteq} \\
  &\Concr_{L_1, \sqsubseteq_1, L_2, \sqsubseteq_2, \gamma} :\equiv\; \forall \overline{x}, \overline{y} \in L_2.\; \overline{x}\sqsubseteq_2 \overline{y}\Rightarrow \gamma(\overline{x})\sqsubseteq_1 \gamma(\overline{y})\\
  &\mathsf{AscCn}_{L, \sqsubseteq}(s) :\equiv\; \forall n, m \in \N.\; \bigl( n\leq m \Rightarrow s(n) \sqsubseteq s(m)\bigr) \\
  &\mathsf{Sup}_{L, \sqsubseteq}(p, s) :\equiv\; \bigl(\forall n \in \N.\; s(n) \sqsubseteq p\bigr) \wedge \forall q \in L. \Bigl( \bigl( \forall n \in \N.\; s(n) \sqsubseteq q \bigr) \Rightarrow p \sqsubseteq q \Bigr) \\
  &\Cpo_{L, \sqsubseteq} :\equiv\; \Poset_{L, \sqsubseteq} \wedge \forall s \in \N \rightarrow L.\; \bigl(\AscCn_{L, \sqsubseteq}(s) \Rightarrow \exists p \in L.\; \mathsf{Sup}_{L, \sqsubseteq}(p, s)\bigr) \\
  &\Monotone_{L_1, \sqsubseteq_1, L_2, \sqsubseteq_2}(f) :\equiv\; \forall x, y \in L_1 .\; x \sqsubseteq_1 y \Rightarrow f(x) \sqsubseteq_2 f(y) \\
  &\Conti_{L_1, \sqsubseteq_1, L_2, \sqsubseteq_2}(f) :\equiv \forall s \in \N \rightarrow L_1.\; \forall p \in L_1.\; \\
  &\hspace{2.5cm}\Bigl(\bigl(\AscCn_{L_1, \sqsubseteq_1}(s) \wedge \mathsf{Sup}_{L_1, \sqsubseteq_1}(p, s)\bigr) \Rightarrow \mathsf{Sup}_{L_2, \sqsubseteq_2}\bigl(f(p), f\circ s\bigr)\Bigr) \\
  &\Basis_{L, \sqsubseteq}(\basis, f) :\equiv\; \basis \sqsubseteq f(\basis) \\
  &\Cover_{L, \sqsubseteq, \nabla} :\equiv\; \forall x, y \in L. \; (x\sqsubseteq x \nabla y) \wedge y \sqsubseteq x \nabla y) \\
 &\Term_{L, \sqsubseteq, \nabla} :\equiv\; \forall x \in \N \rightarrow L .\; \AscCn(x)\Rightarrow \\
  &\quad \biggl( \forall y \in \N \rightarrow L .\;  \Bigl( \bigl( y(0) = x(0) \wedge  \forall n \in \N.\; y(n+1) = y(n) \nabla x(n+1) \bigr) \\
  &\hspace{3cm} \Rightarrow \exists k \in \N. \; y(k) = y(k+1)\Bigr) \biggr) \\
  &\Widen_{L, \sqsubseteq, \nabla} :\equiv\; \Cover_{L, \sqsubseteq, \nabla} \wedge \Term_{L, \sqsubseteq, \nabla} \\
  &\WidenSeq_{L, \sqsubseteq, \nabla}(X, \basis, F) :\equiv\; \\
  &\qquad X(0) = \basis \; \wedge \; \forall n \in \N.\; X(n+1) = X(n) \nabla F(X(n)).
 \end{align*}
\end{mydef}

\subsection{Proof of Prop.~\ref{prop:hyperconvexpoly}}
\begin{myproof}
The constraint system $C$ for a (standard) convex polyhedron $P \in
 \CP_n$ can be expressed by a pair $(\mathbf{A},\mathbf{b})$ of an $m
 \times n$-matrix $\mathbf{A}$ and an $m$-vector $\mathbf{b}$, where $m$
 is the number of linear inequalities in $C$.
 The same applies to a (nonstandard) convex polyhedron
 $P\in\CP^{\hyper{\R}}_{n}$.
 For each of $X\in\{\R,\hyper{\R}\}$,  let us denote,
 by $\Constr_{X, m, n}$, the set of all
 convex polyhedra  over $X^{n}$ that can be expressed with $m$ linear inequalities.


Then $\CP_n = \bigcup_{m\in\N}\Constr_{\R,m,n}$ (with $\bigcup_{m\in\N}$
 expressed using an existential quantifier $\exists m\in\N$) is a valid $\LR$-sentence by Def.~\ref{def:convexPoly}.
By the transfer principle (Lem.~\ref{lemma:transferPrinciple}), we have a valid $\LsR$-sentence $\hyper{(\CP_n)} = \bigcup_{m\in\hyper\N}\Constr_{\hyper\R,m,n}$.
It has as its subset the set
 $\CP_{n}^{\hyper{\R}}=\bigcup_{m\in\N}\Constr_{\hyper\R,m,n}$
since $\N\subseteq\hyper{\N}$. This proves the claim.
\myqed
\end{myproof}

\subsection{Proof of Thm.~\ref{thm:hyperconcretization}}
\begin{myproof}
 Let $L, \overline{L} \in \U$ be sets, $\sqsubseteq \in \Pow(L \times L)$ and $\mathrel{\overline{\sqsubseteq}} \in \Pow(\overline{L} \times \overline{L})$ be binary relations on $L$ and $\overline{L}$ respectively, $\alpha : L \rightarrow \overline{L}$ and $\gamma : \overline{L} \rightarrow L$ be functions.
 Then, the following $\LU$-sentence is valid (because it is equivalent to Prop.~\ref{prop:concretization}):
 \begin{align*}
  & \forall F \in L \rightarrow L. \; \forall \overline{F} \in \overline{L} \rightarrow \overline{L}. \; \forall \basis \in L. \; 
  \forall \overline{x} \in \overline{L}. \\ 
  &\Bigl( \Cpo_{L, \sqsubseteq} \wedge \Preord_{\overline{L}, \mathrel{\overline{\sqsubseteq}}} \wedge \Conti_{L, \sqsubseteq, L, \sqsubseteq}(F) \wedge \Monotone_{\overline{L}, \mathrel{\overline{\sqsubseteq}}, \overline{L}, \mathrel{\overline{\sqsubseteq}}}(\overline{F})  \wedge \Concr_{L, \sqsubseteq, \overline{L}, \mathrel{\overline{\sqsubseteq}}, \gamma}  \\
  &\wedge\; 
F\circ\gamma\sqsubseteq \gamma\circ \overline{F} \wedge \basis \sqsubseteq F(\basis) \wedge 
\basis\sqsubseteq\gamma(\overline{x}) \wedge \overline{F}(\overline{x}) \mathrel{\overline{\sqsubseteq}} \overline{x} \\
  &\Rightarrow \lfp_{\basis}F \sqsubseteq \gamma(\overline{x})\Bigr).
 \end{align*}
By applying Lem.~\ref{lemma:transferPrinciple} to this $\LU$-sentence, we have the following valid $\LUpr$-sentence:
 \begin{align*}
  & \forall F \in \hyper{(L \rightarrow L)}. \; \forall \overline{F} \in \hyper{(\overline{L} \rightarrow \overline{L})}. \; \forall \basis \in \hyper{L}. \; 
  \forall \overline{x} \in \hyper{\overline{L}}.\\ 
  &  \Bigl(\hyper\Cpo_{L, \sqsubseteq} \wedge \hyper\Preord_{\overline{L}, \mathrel{\overline{\sqsubseteq}}} \wedge \hyper\Conti_{L, \sqsubseteq, L, \sqsubseteq}(F) \wedge \hyper\Monotone_{\overline{L}, \mathrel{\overline{\sqsubseteq}}, \overline{L}, \mathrel{\overline{\sqsubseteq}}}(\overline{F}) \wedge \hyper{\Concr}_{L, \sqsubseteq, \overline{L}, \mathrel{\overline{\sqsubseteq}}, \gamma} \\
  & \wedge\; F\circ\hyper{\gamma}\mathrel{\hyper{\sqsubseteq}} \hyper{\gamma}\circ \overline{F} \wedge \basis \mathrel{\hyper{\sqsubseteq}} F(\basis) \wedge {\basis} \mathrel{\hyper{\sqsubseteq}} \hyper{\gamma}(\overline{x}) \wedge \overline{F}(\overline{x}) \mathrel{\hyper{\overline{\sqsubseteq}}} \overline{x} \\
  & \Rightarrow \hyper\lfp_{\basis}F \mathrel{\hyper{\sqsubseteq}} \hyper\gamma(\overline{x})\Bigr).
 \end{align*}
This yields the statement of this theorem. 
For example, we can confirm that $\hyper{\Concr}_{L, \sqsubseteq, \overline{L}, \mathrel{\overline{\sqsubseteq}}, \gamma}$ holds from the following hypothesis in the theorem statement:
 $\overline{a}\mathrel{\overline{\sqsubseteq}}\overline{b} \Rightarrow \gamma(\overline{a})\sqsubseteq \gamma(\overline{b})$ for all $\overline{a}, \overline{b}\in\overline{L}$.

\myqed
\end{myproof}

\subsection{Proof of Thm.~\ref{thm:widenwithinf}}
\begin{myproof}
 Let $L \in \U$ be a set, $\sqsubseteq \in \Pow(L \times L)$ be a binary relation on $L$ and $\nabla : L \times L \rightarrow L$ be a function.
 Then, the following $\LU$-sentence is valid (because it is equivalent to Prop.~\ref{prop:widen}):
 \begin{eqnarray*}
  && \forall F \in L \rightarrow L.\; \forall \basis \in L.\; \forall X \in \N \rightarrow L.\; \\
  && \Preord_{L, \sqsubseteq} \wedge \Monotone_{L, \sqsubseteq, L, \sqsubseteq}(F) \wedge \Basis_{L, \sqsubseteq}(\basis, F) \wedge \Widen_{L, \sqsubseteq, \nabla} \\
  && \wedge \WidenSeq_{L, \sqsubseteq, \nabla}(X, \basis, F) \\
  && \Rightarrow \exists i \in \N.\;  \forall j \in \N.\; i \leq j \Rightarrow X(i) = X(j) \\
  && \wedge \forall k \in \N.\; \Bigl( \bigl( \forall l \in \N.\; k \leq l \Rightarrow X(k) = X(l) \bigr) \Rightarrow F \bigl( X(k) \bigr) \sqsubseteq X(k) \Bigr).
 \end{eqnarray*}
By applying Lem.~\ref{lemma:transferPrinciple} to this $\LU$-sentence, we have the following valid $\LU$-sentence:
 \begin{eqnarray*}
  && \forall F \in \hyper{(L \rightarrow L)}.\; \forall \basis \in \hyper{L}.\; \forall X \in \hyper{(\N \rightarrow L)}.\; \\
  && \hyper\Preord_{L, \sqsubseteq} \wedge \hyper\Monotone_{L, \sqsubseteq, L, \sqsubseteq}(F) \wedge \hyper\Basis_{L, \sqsubseteq}(\basis, F) \wedge \hyper\Widen_{L, \sqsubseteq, \nabla} \\
  && \wedge \hyper\WidenSeq_{L, \sqsubseteq, \nabla}(X, \basis, F) \\
  && \Rightarrow \exists i \in \hyper\N.\;  \forall j \in \hyper\N.\; i \leq j \Rightarrow X(i) = X(j) \\
  && \wedge \forall k \in \hyper\N.\; \Bigl( \bigl( \forall l \in \hyper\N.\; k \leq l \Rightarrow X(k) = X(l) \bigr) \Rightarrow F \bigl( X(k) \bigr) \mathrel{\hyper{\sqsubseteq}} X(k) \Bigr)
 \end{eqnarray*}
This yields the statement of this theorem.
Note that the well-definedness of the iteration hyper-sequence (by
 induction on $i\in\hyper{\N}$) is implicit in the above transfer arguments.
\myqed
\end{myproof}

\subsection{Proof of Thm.~\ref{thm:newunifwidenwithinfinitesimal}}

\begin{myproof}
We can characterize uniform widening operators as an $\LU$-sentence as follows (covering condition has been already expressed as an $\LU$-formula in Def.~\ref{def:lunotations}):
\begin{align*}
 &\UnifTerm_{L, \sqsubseteq, \nabla} :\equiv\; \forall x_0 \in L.\; \exists i \in \N. \; \forall x \in \N \rightarrow L .\; (\AscCn(x) \wedge x(0) = x_0) \Rightarrow\\
 &\quad \biggl( \forall y \in \N \rightarrow L .\; \Bigl( \bigl( y(0) = x(0) \wedge  \forall n \in \N.\; y(n+1) = y(n) \nabla x(n+1) \bigr)  \\
 &\quad \Rightarrow \exists j \in \N. \;\bigl( j \leq i \wedge y(j) = y(j+1)\bigr)\Bigr)\biggr) \\
 &\UnifWiden_{L, \sqsubseteq, \nabla} :\equiv\; \Cover_{L, \sqsubseteq, \nabla} \wedge \UnifTerm_{L, \sqsubseteq, \nabla}
\end{align*}
 Let $L \in \U$ be a set, ${\sqsubseteq} \in \Pow(L \times L)$ be a binary relation on $L$ and $\nabla: L\times L \rightarrow L$ be a function.
 Then, we can see directly that the following $\LU$-sentence is valid:
 \begin{align}\label{fml:unif}
  & \forall \basis \in L.\; \exists i \in \N.\; \Psi(\basis)(i), 
\end{align} 
where
\begin{align*}
 &\Psi(\basis)(i) = \\
 &\quad \forall F \in L \rightarrow L.\; \forall X \in \N \rightarrow L.\; \\
 &\quad \Preord_{L, \sqsubseteq} \wedge \Monotone_{L, \sqsubseteq, L, \sqsubseteq}(F) \wedge \Basis_{L, \sqsubseteq}(\basis, F) \wedge \UnifWiden_{L, \sqsubseteq, \nabla}\\
 &\quad \wedge \WidenSeq_{L, \sqsubseteq, \nabla}(X, \basis, F)\\
 &\quad \Rightarrow \forall j \in \N.\; i \leq j \Rightarrow X(i) = X(j) \\
 &\quad \wedge \forall k \in \N.\; \Bigl( \bigl( \forall l \in \N.\; k \leq l \Rightarrow X(k) = X(l) \bigr) \Rightarrow F \bigl( X(k) \bigr) \sqsubseteq X(k) \Bigr).
 \end{align*}

Assume that $\basis \in L$ is given. Then, by the $\LU$-sentence~(\ref{fml:unif}), there exists $i \in \N$ such that $\Psi(\basis)(i)$ holds.
Therefore, by transferring $\Psi(\basis)(i)$, $\hyper\Psi(\basis)(i)$ holds for such $i\in\N$.
 Note that
$\hyper\Psi(\basis)(i)$ is the following $\LUpr$-sentence ($\basis$ and $i$ are dealt with as constants in the following $\LUpr$-sentence because $\basis$ and $i$ are defined outside the $\LUpr$-sentence):
 \begin{eqnarray*}
  && \forall F \in \hyper{(L \rightarrow L)}.\; \forall X \in \hyper{(\N \rightarrow L)}.\; \\
  && \hyper{\Preord}_{L, \sqsubseteq} \wedge \hyper{\Monotone}_{L, \sqsubseteq, L, \sqsubseteq}(F) \wedge \hyper{\Basis}_{L, \sqsubseteq}(\hyper\basis, F) \wedge \hyper{\UnifWiden}_{L, \sqsubseteq, \nabla} \\
  && \wedge \hyper{\WidenSeq}_{L, \sqsubseteq, \nabla}(X, \hyper\basis, F) \\
  && \Rightarrow \forall j \in \hyper{\N}.\; i \leq j \Rightarrow X(i) = X(j) \\
  && \wedge \forall k \in \hyper{\N}.\; \Bigl( \bigl( \forall l \in \hyper{\N}.\; k \leq l \Rightarrow X(k) = X(l) \bigr) \Rightarrow F \bigl( X(k) \bigr) \mathrel{\hyper{\sqsubseteq}} X(k) \Bigr).
 \end{eqnarray*}
This yields Thm.~\ref{thm:newunifwidenwithinfinitesimal}.
\myqed
\end{myproof}

\subsection{Proof of Thm.~\ref{thm:uniformityOfKnownWidening}}
We prove the uniformity and nonuniformity of three 
 widening operators (
 $\nabla_S$, $\nabla_M$, $\nabla_N$) in this order.

       \begin{myproof}
	Let $\langle X_i \rangle_i$ be a iteration sequence defined by $\nabla_{\CP_n}$, a basis $X_0 = \con(C_0)$ and a monotone function $F$.
Let $\langle C_i \rangle_i$ be the sequence of constraint systems that corresponds to $\langle X_i \rangle_i$.
	By definition of $\nabla_{\CP_n}$ and the construction of $\langle X_i \rangle_i$, regardless of the function $F$, $C_{i+1}\subseteq C_i$ for all $i \in \N$.
	Thus for any basis $X_0 = \con(C_0)$ and monotone function $F$, we can reach a prefixed point by iterating the widening operator at most $\#(C_0)$ times and this means the widening operator $\nabla_{\CP_n}$ is uniform.
\myqed
       \end{myproof}

        \begin{myproof}
	 The constraints in $M$ may be added in the iteration sequence, but by the definition of the standard widening $\nabla_S$, a constraint in $M$ will never appear once it is violated.
	 Therefore the number of steps for an iteration sequence to converge is at most $\#(M)$ larger than the case of standard widening.
\myqed
	\end{myproof}

       \begin{myproof}
	Assume that $P_1 = \con\{0\leq x\leq 1, 0\leq y\leq 1, z=0\} \in \CP_3$, $P_2 \in \CP_3$ includes $P_1$ and the linear equation ``$z=0$'' is not included in $C_2$.
	Then $P_1 \curvearrowright_N P_2$ holds because $\#eq(C_1) > \#eq(C_2)$.
	The maximum number of steps for an iteration sequence starting from $P_2$ to converge is $\#C_2$.
	This is not limited uniformly because you can define $P_2$ such that $\#C_2$ is as large as you like.
\myqed
       \end{myproof}




\begin{mydef}[Galois connection]\label{def:galois}
 Let $(L, \sqsubseteq)$ and $(\overline{L},
 \mathrel{\overline{\sqsubseteq}})$ be posets, and
 $\alpha: L \rightarrow \overline{L}$ and $\gamma: \overline{L} \rightarrow L$ be functions.
 A tuple $\left((L, \sqsubseteq), (\overline{L},
 \mathrel{\overline{\sqsubseteq}}), \alpha, \gamma\right)$ is said to be
 a \emph{Galois connection} if: for each $x\in L$ and
 $\overline{y}\in\overline{L}$, we have 
 $\alpha{x} \mathrel{\overline{\sqsubseteq}} \overline{y} \Leftrightarrow x \sqsubseteq \gamma\overline{y}$.
 This fact is denoted by  $\Galois{L}{\overline{L}}{\gamma}{\alpha}$;
 and we call $L$ a \emph{concrete domain}, $\overline L$ an
 \emph{abstract domain}, $\alpha$ an \emph{abstraction function} and
 $\gamma$ a \emph{concretization function}.
\end{mydef}
\begin{myprop}\label{prop:galoisExtendsToFuncSp}
A Galois connection $\Galois{(L, \sqsubseteq)}{(\overline{L},
 \overline\sqsubseteq)}{\gamma}{\alpha}$ extends to monotone
 endofunctions. Concretely, it yields a Galois connection
 $\Galois{(L\underset{\mathrm{mono.}}{\longrightarrow} L)}
 {(\overline{L}\underset{\mathrm{mono.}}{\longrightarrow}\overline{L})}{\vec\gamma}{\vec\alpha}$
where $L\underset{\mathrm{mono.}}\longrightarrow L$ and
 $\overline{L}\underset{\mathrm{mono.}}\longrightarrow\overline{L}$ are
 the posets of monotone functions ordered by the pointwise extension of
 $\sqsubseteq$ and $\overline\sqsubseteq$. The functions $\vec\gamma$
 and $\vec\alpha$ here are defined by: $\vec\gamma(f) = \gamma \circ f \circ \alpha$ and $\vec\alpha(f) = \alpha \circ f \circ \gamma$, respectively.
\myqed
\end{myprop}

\begin{myprop}\label{prop:galois}
In the above setting, assume further that:  $\overline{F}: \overline{L} \rightarrow
 \overline{L}$ be a monotone function such that
 $F\sqsubseteq\vec\gamma(\overline{F})$; and that 
$\overline{x}\in\overline{L}$ is a prefixed point of  $\overline{F}$ 
 (i.e.\ $\overline{F}(\overline{x})\mathrel{\overline\sqsubseteq}\overline{x}$)
 such that $\alpha(\basis)\mathrel{\overline\sqsubseteq}\overline{x}$.

Then $\overline{x}$ over-approximates $\lfp_{\basis}F$, that is,
$\lfp_{\basis}F\sqsubseteq \gamma(\overline{x})$.
\myqed
\end{myprop}

\begin{mydef}[hyper-Galois connection]
\label{def:hyperGalois}
 A \emph{hyper-Galois connection}, denoted by
 $\Galois{\hyper{L}}{\hyper{\overline{L}}}{\hyper\gamma}{\hyper\alpha}$,
 is a quintuple
 $\bigl(\hyper{L},\hyper{\overline{L}},\hyper\alpha,\hyper\gamma\bigr)$
 of: the $*$-transform of a poset $L$; that of a poset $\overline{L}$;
 the $*$-transform $\hyper{\alpha}\colon \hyper{L}\to\hyper{\overline{L}}$ of a
 function
 $\alpha\colon L\to\overline{L}$; and  the $*$-transform
 $\hyper{\gamma}\colon
 \hyper{\overline{L}}\to\hyper{L}$ of $\gamma$. We require that the data
 $(L,\overline{L}, \alpha,\gamma)$ forms a Galois connection (Def.~\ref{def:galois}).
\end{mydef}
The above $\hyper\alpha$ is an internal
function (i.e.\ $\hyper\alpha\in\hyper{(L\to\overline{L})}$);
see Appendix~\ref{appendix:NSAPrimer} for details.
The notion of \emph{$*$-continuous function} $f'\colon
\hyper{L}\to\hyper{L}$ is defined analogously, namely that $f'$ is 
the $*$-transform of some continuous function $f\colon L\to L$. See
Appendix~\ref{appendix:domainTheoryTransferred}. 

Here is the counterpart of Prop.~\ref{prop:galois}. As
announced, it only requires the cpo structure of $L$ (not of $\hyper{L}$) and the $*$-continuity of $F$.

\begin{mythm}
\label{thm:newgaloiswithinf}
 Let $(L, \sqsubseteq)$ be a cpo, $(\overline L, \overline\sqsubseteq)$
 be a poset such that $\Galois{L}{\overline{L}}{\gamma}{\alpha}$, and
 consider the induced hyper-Galois connection 
 $\Galois{\hyper{L}}{\hyper{\overline{L}}}{\hyper\gamma}{\hyper\alpha}$.
 Let $F: \hyper L \rightarrow \hyper L$ 
 be a $*$-continuous function;  
  $\basis\in\hyper{L}$ be such that
$\basis \mathrel{\hyper\sqsubseteq} F(\basis)$, and 
 $\overline{F}: \hyper{\overline{L}} \rightarrow \hyper{\overline{L}}$ 
 be an internal function
 that is monotone with respect to $\hyper{\overline{\sqsubseteq}}$.
 Assume that $F\mathrel{\hyper{\sqsubseteq}}(\vec{\hyper{\gamma}})(\overline{F})$;
 and that $\overline{x}\in\hyper{\overline{L}}$ is a prefixed
 point of $\overline{F}$, i.e.\
 $\overline{F}(\overline{x})\mathrel{\hyper{\overline{\sqsubseteq}}}\overline{x}$.

Then $\overline{x}$ over-approximates $\lfp_{\basis}F$, that is, 
 $\lfp_{\basis}F\mathrel{\hyper{\sqsubseteq}}\hyper\gamma(\overline{x})$.
\myqed
\end{mythm}
\end{document}